\documentclass[12pt]{article}
\usepackage{graphicx}
\usepackage{psfrag}

\newcommand{\Ree}{\Re\mathrm{e}}
\newcommand{\Li}{\mbox{Li}_2}
\newcommand{\dd}{\mbox{d}}
\newcommand{\eps}{\varepsilon}

\title{Monte-Carlo generator for $e^+e^-$ annihilation 
into lepton and hadron pairs with precise radiative corrections}

\author{
A.B.~Arbuzov$^1$,
G.V.~Fedotovich$^2$,
F.V.~Ignatov$^2$, \\
E.A.~Kuraev$^1$,
A.L.~Sibidanov$^2$
}
\date{}

\begin{document}

\maketitle 
\begin{center}
{$^1$ \it Bogoliubov Laboratory of Theoretical Physics, JINR, \\
Dubna, 141980, Russia} \\[.2cm]
{$^2$ \it Budker Institute for Nuclear Physics, \\
Prospect Lavrent'eva, 11, Novosibirsk, 630090, Russia}
\end{center}

\begin{abstract}
Recently, various cross sections of $e^+e^-$ annihilation into hadrons 
were accurately measured in the energy range from 0.37 to 1.39 GeV 
with the CMD-2 detector at the VEPP-2M collider. In the $\pi^+\pi^-$ 
channel a systematic uncertainty of $0.6\%$ has been achieved. 
A Monte-Carlo Generator  Photon Jets (MCGPJ) was developed to simulate 
events of the Bhabha scattering as well as production of two charged 
pions, kaons and muons. Based on the formalism of Structure Functions, 
the leading logarithmic contributions related to the emission of photon 
jets in the collinear region are incorporated into the MC generator.
Radiative corrections (RC) in the first order of $\alpha$ are accounted for
exactly. The theoretical precision of the cross sections with RC   
is estimated to be better than $0.2\%$. 
Numerous tests of the program as well as comparison with other 
MC generators and CMD-2 experimental data are presented. 
\end{abstract}
\section{Introduction}
The cross sections of $e^+e^-$ annihilation into 
hadrons are very important in various problems of 
particle physics and, in particular, they are required  
for the evaluation of the hadronic contribution to 
the anomalous magnetic moment of muon, $a_{\mu} = (g-2)_{\mu}/2$. 
The recent measurement of $a_{\mu}$ at BNL~\cite{e821} led to a new world 
average differing by 2.7 standard deviations 
from its theoretical evaluation. One of the main ingredients in the 
theoretical prediction for $a_{\mu}$ is the hadronic contribution  
related via a dispersion integral to the cross 
section of $e^+e^-$ annihilation into hadrons. In the case of  
$a_{\mu}^{had}$, the VEPP-2M energy range gives the major contribution both to 
the hadronic vacuum polarization contribution itself and to its 
uncertainty~\cite{jeger, cmd2}. This uncertainty is dominated by systematic  
errors of the experimental values of $R(s)$ which are used as an input to 
the integral with the proper kernel function
~\cite{integr}:    
$$a^{had}_{\mu} = \biggl(\frac{\alpha m_{\mu}}{3\pi}\biggr)^2 
\int\limits_{4m^2_{\pi}}^{\infty}\frac{R(s)K(s)}{s^2}\dd s.$$
The quantity $R(s)$ is defined as  
$ R(s) = \sigma (e^+e^- \to {\mathrm{hadrons}})/\sigma (e^+e^- \to \mu^+\mu^-)$
and at high energies can be calculated within the QCD framework whereas 
at low energies the experimental data are required.  
A numerical computation of this integral can be found elsewhere\cite{jeger} 
and in relative unities its evaluation yields the result $\sim$ 70~ppm.

The goal of the new BNL experiment~\cite{e969} is to measure the anomalous 
magnetic moment of muon with the relative accuracy $\sim$ 0.25~ppm.  
To reduce the current systematic error of the hadronic contribution to 
$a^{had}_{\mu}$ at least to the same level, the theoretical precision of 
the cross sections with radiative corrections (RC) should be better 
than $0.3\%$ as it follows from a simple estimation: 
70~ppm $\times 0.3\% \sim 0.2$~ppm. This short observation shows why
the knowledge of the cross sections $e^+e^-$ annihilation into hadrons
with high precision is extremely important.  

The detection efficiency, background conditions and criteria of event 
selection from the raw data differ for specific $e^+e^-$ annihilation modes.   
Therefore, expressions for the cross sections with RC on which 
the MC generator is based, should have a completely differential form with 
respect to the kinematic variables of final particles.  
In this case the influence of the selection criteria as well as the 
trigger efficiency and many other specific resolutions of the 
detector can be naturally incorporated in a MC generator. 

During the last thirty years considerable efforts were devoted to 
elucidate theoretical understanding of the accuracy of cross 
sections with RC, particularly in the case of $e^+e^-$ and 
$\pi^+\pi^-$ pair production at low energies.  
The radiatively corrected cross sections for annihilation channels 
with an accuracy of about $0.1\%$ were obtained in~\cite{fadkur}.
Unfortunately, expressions for these cross sections do not contain the angular 
distributions for the emitted photons and, as a result, it is not possible
to reconstruct the kinematics of the final particles correctly.  
On the other hand, the differential cross sections were obtained 
in~\cite{berends}, but their relative accuracy is about $1\%$, since only 
${\mathcal O}(\alpha)$ QED corrections were taken into account. 

The work~\cite{arkurlept} is based in part on a combination of the 
approaches~\cite{fadkur, berends} mentioned above. 
To achieve the accuracy 
$\sim 0.2\%$, higher order radiative corrections were taken into account 
    by means of the Structure Function (SF) formalism~\cite{fadkur}.
    It involves a convolution of the {\em boosted} Born cross section 
    with the electron (positron) SF, which   
    describes the leading effects due to emission of photons 
in the collinear region as well as radiation of $e^+e^-$ pairs. 
These enhanced contributions are proportional 
to $(\alpha/\pi)^n \ln^n(s/m^2_{e})$, $n = 1,2,...$ and are referred 
to as the leading ones. Moreover, in the 
{\em smoothed} representations of the 
SF~\cite{fadkur} a certain part of these corrections is exponentiated 
and evaluated in all powers of n. The non-leading contributions 
proportional to $(\alpha/\pi)$ are incorporated exactly 
according to~\cite{fadkur} by means of a so-called {\em K-factor}. The 
next-to-leading contributions of the second order 
$(\alpha/\pi)^2  \ln(s/m^2_{e}) \sim 0.01\%$ are fortunately small and can 
be omitted, keeping in mind the intended precision tag 0.2\%.
      
The vacuum polarization effects in the photon propagator 
are treated as in~\cite{arkurlept} for the lepton channels. 
These effects are not included in RC for the hadronic modes according 
to the generally accepted agreement~\cite{tsai}. 
The emission of one hard photon at a large angle is described by a 
differential formula, which allows to take into account specific 
experimental conditions and cuts. Based on numerical calculations, 
which will be given below, it was established that the 
$\mathcal O(\alpha)$ radiative corrections together with photon jet 
radiation in a collinear region are sufficient to achieve   
a relative precision of the cross sections about $\sim$ 0.2\%.  

The purpose of this paper is to describe the Monte-Carlo Generator Photon 
Jets (MCGPJ) which simulates processes $e^+e^- \to e^+e^-$, $\mu^+\mu^-$, 
$\pi^+\pi^-$, $K^+K^-$ and $K_{L}K_{S}$. This generator was used while 
CMD-2 experimental data were processed. 
The MCGPJ code has a modular structure that simplifies the implementation 
of additional hadronic modes as well as the replacement of
matrix elements of the current cross sections by a new one. The effects 
of the final state radiation (FSR) for the channels 
$\mu^+\mu^-, \pi^+\pi^-, K^+K^- $ have been 
incorporated into the program. The pions were assumed to be point-like objects,
and the scalar QED was applied to calculate virtual, soft and hard photon
emission by charged pions (kaons).    

The relevant formulae for the cross sections with RC of order $\alpha$  
are collected here from many other papers.
It is done specially to have a possibility of quantifying  
the difference between cross sections due to radiation of one photon 
and photon jets in the collinear region. On the other hand, some expressions 
for RC are revisited in the present paper and explicit analytical formulae 
will be presented in the form convenient for 
the MC generator construction.

\section{Monte-Carlo generator for events of Bhabha \\
scattering at large angles}
The {\em boosted\/} Born cross section of the  process
$ e^{-}(z_{1}p_{-})+e^{+}(z_{2}p_{+})\to e^{-}(p_{1})+e^{+}(p_{2}),$
corrected for vacuum polarization factors in the $s$ and $t$ channels, 
when initial particles lose some energy by radiation of 
photon jets in the collinear region, has the following 
form~\cite{arkurlept} in the c.m. frame:  
\begin{eqnarray}       
  & & \frac{\dd\tilde{\sigma}^{e^+e^- \to e^+e^-}_{0}(z_{1},z_{2})}{
\dd\Omega_{1}} = \frac{4z_{1}z_{2}\alpha^2}{a^2\tilde{s}} 
\biggl(\frac{\tilde{s}^2 + 
\tilde{u}^2}{2\tilde{t}^2 \vert 1 - \Pi(\tilde{t})\vert^2} + 
\frac{\tilde{t}^2 + 
\tilde{u}^2}{2\tilde{s}^2 \vert 1 - \Pi(\tilde{s})\vert^2} \nonumber \\
  & & \qquad + \Ree \biggl\{\frac{\tilde{u}^2}{\tilde{s}\tilde{t}}
\frac{1}{(1 - \Pi(\tilde{s}))(1 - \Pi(\tilde{t}))}\biggr\} \biggr), 
\label{eq:bornshift} 
\end{eqnarray}
where $z_{1}$ and $z_{2}$ are the electron and positron fraction energies 
after radiation of photon jets $(z_{1,2} = 
\varepsilon_{1,2}/\varepsilon_{beam})$, 
$\Pi(\tilde s)$ and $\Pi(\tilde t)$ are the vacuum polarization 
operators in photon propagators in the $s$ and $t$ channels,  
respectively. The Mandelstam variables in the Lab and c.m.s. are 
defined as usual:
$s = 2p_-p_+,$ 
$t = -2p_-p_1,$ 
$u = -2p_-p_2, $
$\tilde{s} = sz_{1}z_{2},$
$\tilde{t} = -sz_{1}Y_{1}(1-c_{1})/2,$
$\tilde{u} = -sz_{2}Y_{1}(1+c_{1})/2,$ 
$s_1= 2p_1p_2,$   
$t_1=-2p_+ p_2,$   
$u_1 = -2p_+ p_2,$
$c_{1}=\cos{\theta_{1}}$, where $\theta_{1}$ is a polar angle of the 
final electron with respect to the electron beam direction, 
$Y_{1}$ and $Y_{2}$ are the relative energies of final $e^-$ and $ e^+$.
The energy-momentum conservation law allows to reconstruct the 
kinematics of final particles 
and to find these energies and a positron polar angle $\theta_{2}$:
$z_{1} + z_{2} = Y_{1} +Y_{2}$ - energy conservation; 
$z_{1} - z_{2} = Y_{1}\cos{\theta_{1}} +Y_{2}\cos{\theta_{2}}$ 
- momentum conservation along the Z-axis;
$Y_{1}\sin{\theta_{1}} = Y_{2}\sin{\theta_{2}}$ - momentum 
conservation in the plane perpendicular to the Z-axis.
From these equations one can find that 
\begin{eqnarray}
& & Y_{1} = \frac{2z_{1}z_{2}}{a}, \hspace{3 cm} Y_{2} = \frac
    {(z_{1}^2 + z_{2}^2) - (z_{1}^2 - z_{2}^2)c_{1}}{a}, 
\nonumber \\ 
& & c_{2} = \frac{(z_{1}^2 - z_{2}^2) - (z_{1}^2 + z_{2}^2)c_{1}}
    {(z_{1}^2 + z_{2}^2) - (z_{1}^2 - z_{2}^2)c_{1}},  \hspace{0.2 cm}
{\rm where} \hspace{0.2 cm} a = z_{1} + z_{2} - (z_{1} - z_{2})c_{1}.
\end{eqnarray} 

The expression for the differential cross section with one
photon emission in the reaction 
$ e^{-}(p_{-})+e^{+}(p_{+})\to e^{-}(p_{1})+e^{+}(p_{2})+\gamma(k),$
was obtained in~\cite{berends} (see also references therein) and reads 
\begin{equation}
    \dd\sigma^{e^+e^- \to e^+e^-\gamma}_{\mathrm{hard}} = 
  \frac{\alpha^3}{2\pi^{2}s}
  R^{e^+e^- \to e^+e^-\gamma}_{hard}
    \frac{\dd^3 p_{1}}{\varepsilon_{1}}
    \frac{\dd^3 p_{2}}{\varepsilon_{2}} 
    \frac{\dd^3 k}{\omega} 
\delta^{(4)}(p_{-} + p_{+} - p_{1} - p_{2} - k),
\label{eq:hard}
\end{equation}
where $\varepsilon_{1}$, $\varepsilon_{2}$, and $\omega$
are the energies of the final state electron, positron and photon, 
respectively; $\delta$-function provides the energy-momentum conservation.

The expression for the quantity $R^{e^+e^- \to e^+e^-\gamma}_{hard}$ which 
contains the vacuum polarization effects in photon propagators was 
derived in~\cite{arkurlept}:  
\begin{eqnarray}
  & & R^{e^+e^- \to e^+e^-\gamma}_{hard} = \frac{(WT)_{\Pi}}{4} 
  \label{eq:onephotvp} 
  \nonumber \\ 
  & & - \frac{m^{2}_e}{\chi'_{+}\!\!{}^2}
  \biggl(\frac{s^2 + (s + t)^2}{2t^2
    (1 - \Pi(t))^2}
  + \frac{t^2 + (s + t)^2}{2s^2
    \vert 1 - \Pi(s)\vert^2} 
  + \Ree\biggl\{\frac{(s + t)^2}{st 
    (1 - \Pi(s))(1 - \Pi(t))}\biggr\}\biggr) 
  \nonumber \\ 
  & & - \frac{m^{2}_e}{\chi'_{-}\!\!{}^2}
  \biggl(\frac{s^2 + (s + t_1)^2}{2t_1^2
    (1 - \Pi(t_1))^2}
  + \frac{t_1^2 + (s + t_1)^2}{2s^2
    \vert 1 - \Pi(s)\vert^2} 
  +  \Ree\biggl\{\frac{(s + t_1)^2}{st_1 
    (1 - \Pi(s))(1 - \Pi(t_1))}\biggr\}\biggr)
  \nonumber \\ 
  & & - \frac{m^{2}_e}{\chi_{+}^2}
  \biggl(\frac{s_1^2 + (s_1 + t)^2}{2t^2
    ( 1 - \Pi(t))^2} 
  + \frac{t^2 + (s_1 + t)^2}{2s_1^2
    \vert 1 - \Pi(s_1)\vert^2}   
  +  \Ree\biggl\{ \frac{(s_1 + t)^2}{s_1t 
    (1 - \Pi(s_1))(1 - \Pi(t))}\biggr\}\biggr)
  \nonumber \\ & & 
  - \frac{m^{2}_e}{\chi_{-}^2}
  \biggl(\frac{s_1^2 + (s_1 + 
    t_1)^2}{2t_1^2
    ( 1 - \Pi(t_1))^2} 
  + \frac{t_1^2 + (s_1 + t_1)^2}{2s_1^2
    \vert 1 - \Pi(s_1)\vert^2} 
  +  \Ree\biggl\{\frac{(s_1 + t_1)^2}{s_1t_1 
    (1 - \Pi(s_1))(1 - \Pi(t_1))}\biggr\}\biggr),   
\end{eqnarray}
where $\chi_{\pm}=kp_{\pm}$ and $\chi_{\pm}'=kp_{1,2}.$
The quantity $(WT)_{\Pi}$ describes the process with one hard photon emission 
and gives the dominant contribution outside the collinear 
region~\cite{arkurlept}:
\begin{eqnarray}
& & (WT)_{\Pi}=\frac{SS}{\vert 1-\Pi(s)\vert ^2s\chi'_{-}\chi'_{+}} 
+ \frac{S_{1}S_{1}}{\vert 1-\Pi(s_{1})\vert ^2 s_{1}\chi_{-}\chi_{+}} 
- \frac{TT}{\vert 1 - \Pi(t)\vert ^2 t\chi_{+}\chi'_{+}} 
\nonumber \\ 
& & - \frac{T_{1}T_{1}}{\vert 1 - \Pi(t_1)\vert ^2 
t_{1}\chi_{-}\chi'_{-}} + 
\Ree \biggl[\frac{TT_{1}}{(1-\Pi(t))(1-\Pi(t_1))
tt_{1}\chi_{-}\chi'_{-}\chi_{+}\chi'_{+}}\nonumber \\ 
& & - \frac{SS_{1}}{(1 - \Pi(s))(1 - \Pi(s_1))^{*}
ss_{1}\chi_{-}\chi'_{-}\chi_{+}\chi'_{+}} 
+ \frac{TS}{(1 - \Pi(t))(1 - \Pi(s))
ts\chi'_{-}\chi_{+}\chi'_{+}} \nonumber \\ 
& & + \frac{T_{1}S_{1}}{(1 - \Pi(t_{1}))
(1 - \Pi(s_1))t_{1}s_{1}\chi_{-}\chi'_{-}\chi_{+}}  
- \frac{T_{1}S}{(1 - \Pi(t_{1}))
(1 - \Pi(s))t_{1}s\chi_{-}\chi'_{-}\chi'_{+}} \nonumber \\ 
& &- \frac{TS_{1}}{(1 - \Pi(\tilde{t}))(1 - \Pi(\tilde{s}_1))
ts_{1}\chi_{-}\chi_{+}\chi'_{+}} \biggr],  
\end{eqnarray}
where the following notations were used:
\begin{eqnarray}
& & SS = S_{1}S_{1} = t^2 + t_{1}^2 + u^2 + u_{1}^2, 
\nonumber \\ 
& & TT = T_{1}T_{1} = s^2 + s_{1}^2 + u^2 + u_{1}^2, 
\nonumber \\
& & SS_{1} = (t^2 + t_{1}^2 + u^2 + u_{1}^2) 
\times(t\chi_{+}\chi'_{+} + t_{1}\chi_{-}\chi'_{-} - 
u\chi_{+}\chi'_{-} - u_{1}\chi_{-}\chi'_{+}), 
\nonumber \\
& & TT_{1} = (s^2 + s_{1}^2 + u^2 + u_{1}^2)   
\times(u\chi_{+}\chi'_{-} + u_{1}\chi_{-}\chi'_{+} + 
s\chi_{-}'\chi_{+}' + s_{1}\chi_{-}\chi_{+}), 
\nonumber \\
& & TS = -\frac{1}{2}(u^2 + u_{1}^2)(s(t + s_{1}) + t(s + t_{1}) - uu_{1}), 
\nonumber \\
& & TS_{1} = -\frac{1}{2}(u^2+u_{1}^2)(t(s_{1}+t_{1})+s_{1}(s+t)-uu_{1}), 
\nonumber \\
& & T_{1}S = \frac{1}{2}(u^2 + u_{1}^2)(t_{1}(s + t) + s(s_{1} + 
t_{1}) -  uu_{1}), 
\nonumber \\
& & T_{1}S_{1} = \frac{1}{2}(u^2 + u_{1}^2)(s_{1}(s + t_{1}) + 
t_{1}(s_{1} + t) - uu_{1}). 
\end{eqnarray}  

\begin{figure}[htbp]
  \begin{minipage}[t]{0.9\textwidth}
    \includegraphics[width=0.9701\linewidth]{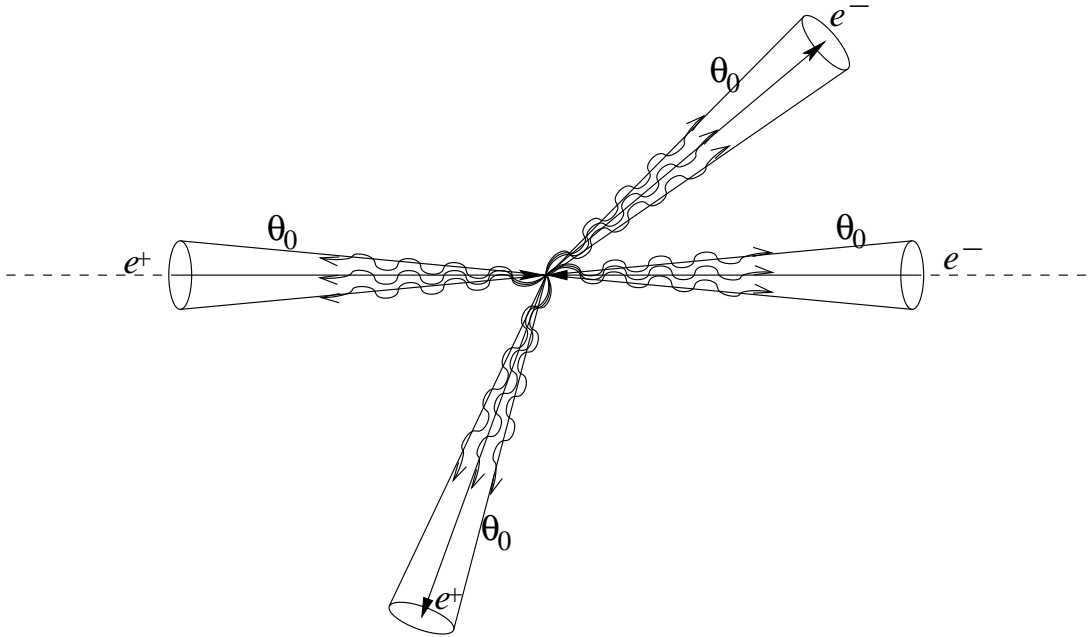}
    \caption{\label{fig:coll} Photon jets are inside four narrow cones 
with an opening angle 2$\theta_{0}.$}
  \end{minipage}
\end{figure}

The main contribution to the cross section due to photon radiation   
comes from the collinear region, where the cross section exhibits 
very steep behavior~\cite{BFKh}.
The collinear region is a part of the angular phase-space with 
four narrow cones (Fig.~\ref{fig:coll}) surrounding the   
  directions of motion of the initial and final particles. 
  The emitted photon should be
  inside these cones with an angle $2\theta_0$. 
  The angle $\theta_0$ should obey to the follow restrictions, $ 1/\gamma \ll
  \theta_0 \ll 1,$ where $\gamma = \varepsilon/m_{e}$. 
It serves as an auxiliary parameter and usually its value is taken 
at about $\sim 1/\sqrt{\gamma}$. The cross section integrated inside 
these narrow cones according to~\cite{arkurlept} is: 
\begin{eqnarray}
& & \frac{\dd\sigma_{\mathrm{coll}}^{e^+e^- \to e^+e^-\gamma}}
            {\dd\Omega_1}=\frac{\alpha}{\pi}
      \int\limits_{\Delta}^{1}\frac{\dd x}{x}\; \Biggl\{
      2\,\frac{\dd\tilde{\sigma}^{e^+e^- \to e^+e^-}_{0}(1,1)}{\dd\Omega_1} 
\left[\left(z+\frac{x^2}{2}\right)
\left(L - 1 + \ln\frac{\theta_0^2 z^2}{4}\right) + \frac{x^2}{2}\right]
\nonumber \\ 
& & + \left[ \frac{\dd\tilde{\sigma}_{0}^{e^+e^- \to e^+e^-}(z,1)}{\dd\Omega_1}
	+ \frac{\dd\tilde{\sigma}_{0}^{e^+e^- \to 
e^+e^-}(1,z)}{\dd\Omega_1} \right]
\biggl[\left(z+\frac{x^2}{2}\right)
	\left(L - 1 + \ln\frac{\theta_0^2}{4}\right) + \frac{x^2}{2}\biggr]
      \Biggr\},
    \label{eq:bhcoll}
\end{eqnarray}
where $L = \ln(s/m_e^2)$, $z = 1 - x$ and the {\em boosted} Born cross 
section is defined in Eq.~(\ref{eq:bornshift}). The auxiliary parameter 
$\Delta$ =$\Delta \varepsilon$/$\varepsilon$ ($\Delta \ll 1$) 
serves as a separator of hard and soft photons, $\varepsilon$ 
is the beam energy. 
The terms proportional to $(\alpha/\pi)(L-1)$ are accounted for in the 
SF~\cite{fadkur} and therefore should be removed from this expression 
to eliminate the double counting.

The remaining four terms can be interpreted as the so-called 
{\em compensators} 
which cancel out the dependence of the total cross section on  
the auxiliary parameter $\theta_0$ when they are summed with the 
cross section Eq.(\ref{eq:hard}) describing one hard photon emission 
outside cones.

Collecting all the discussed above terms into one formula we get the complete 
expression for the {\em master} formula describing the process 
$e^+e^- \to e^+e^- + n\gamma,$ which can be presented as:
\begin{eqnarray}
& & \frac{\dd\sigma^{e^+e^-\to e^+e^- + n\gamma}}{\dd\Omega_1} =  
\int\limits_{0}^{1}\!\!\dd x_1\!\!\int\limits_{0}^{1}\!\!\dd x_2\!\!
 \int\limits^{1}_{0}\!\!\dd x_3\!\!\int\limits^{1}_{0}\!\!\dd x_4
\frac{\dd\tilde{\sigma}_{0}^{e^+e^- \to e^+e^-}(z_1,z_2)}{\dd\Omega_1}
\nonumber \\ & & \quad \times
{\cal D}(z_1,s){\cal D}(z_2,s){\cal D}(z_3,\tilde{s}){\cal D}(z_4,\tilde{s}) 
\left(1+\frac{\alpha}{\pi}\tilde{K}_{SV}\right)\Theta({\mathrm{cuts}})
\nonumber \\ 
& & \quad + \frac{\alpha}{\pi}\int\limits_{\Delta}^{1}\frac{\dd x_{1}}{x_{1}}
      \biggl[\bigl(z_{1} + \frac{x_{1}^2}{2}\bigr) 
	\ln\frac{\theta_{0}^{2}}{4} + \frac{x_{1}^2}{2}\biggr] 
      \frac{\dd\tilde{\sigma}_{0}^{e^+e^- \to e^+e^-}(z_{1},1)}
{\dd\Omega_{1}}\Theta({\mathrm{cuts}})
\nonumber\\
& & \quad + \frac{\alpha}{\pi}\int\limits_{\Delta}^{1}\frac{\dd x_{2}}{x_{2}}
      \biggl[\bigl(z_{2} + \frac{x_{2}^2}{2}\bigr) 
	\ln\frac{\theta_{0}^{2}}{4} + \frac{x_{2}^2}{2}\biggr]
      \frac{\dd\tilde{\sigma}_{0}^{e^+e^- \to e^+e^-}(1,z_{2})}
{\dd\Omega_{1}}\Theta({\mathrm{cuts}})
\nonumber\\
& & \quad +\frac{\alpha}{\pi}\int\limits_{\Delta}^{1}\frac{\dd x_{3}}{x_{3}}
      \biggl[\bigl(z_{3} + \frac{x_{3}^2}{2}\bigr) 
	\ln\frac{\theta_{0}^{2}z_{3}^2}{4} + \frac{x_{3}^2}{2}\biggr]
      \frac{\dd\tilde{\sigma}_{0}^{e^+e^- \to e^+e^-}(1,1)}
{\dd\Omega_{1}}\Theta({\mathrm{cuts}})
\nonumber\\
& & \quad +\frac{\alpha}{\pi}\int\limits_{\Delta}^{1}\frac{\dd x_{4}}{x_{4}}
      \biggl[\bigl(z_{4} + \frac{x_{4}^2}{2}\bigr) 
	\ln\frac{\theta_{0}^{2}z_{4}^2}{4} + \frac{x_{4}^2}{2}\biggr]
      \frac{\dd\tilde{\sigma}_{0}^{e^+e^- \to e^+e^-}(1,1)}
{\dd\Omega_{1}}\Theta({\mathrm{cuts}})
\nonumber\\
& & \quad +\frac{4\alpha}{\pi}
\frac{\dd\tilde{\sigma}_{0}^{e^+e^- \to e^+e^-}(1,1)}{\dd\Omega_{1}}
\ln\frac{u}{t}\ln\Delta
+\frac{\alpha^3}{2\pi^2s}\!\!\!\!\int\limits_{k^{0}>\Delta\varepsilon \atop
\theta_{\gamma}>\theta_{0}}\!\!\!\!R^{e^+e^- \to e^+e^-\gamma}_{hard}
\frac{\dd\Gamma}{\dd\Omega_{1}}\Theta({\mathrm{cuts}}),
    \label{eq:master}
\end{eqnarray} 
where $x_{1,2,3,4}$ are the relative energies of 
photon jets emitted along the initial and final particles; 
$z_{1,2,3,4}=1-x_{1,2,3,4}$ are the energy fractions of electrons 
and positrons after radiation of photon jets;
$\Theta({\mathrm{cuts}})$ is a step-function   
equal to 1 (0) if the kinematics variables obey (or not) 
the selection criteria; 
the expression for $\tilde{K}_{SV}(\tilde{\theta}_{1})$ can be found 
in~\cite{berends, arkurlept}. More details concerning the implementation 
of the SF formalism as adopted in the present paper can be found  
in~\cite{fadkur}. 

The SF approach provides essential improvement of accuracy for the 
Bhabha cross section by taking into account radiation of photon jets 
in the collinear region. These improvements as well as others performed 
in~\cite{arkurlept} are summarized below:\\
1. The radiation of photon jets (enhanced contributions) is taken into  
account by means of the SF formalism.\\ 
2. To combine the cross sections, describing radiation of one hard photon 
inside and outside narrow cones, the four {\em compensators} are embedded 
into the {\em master} formula (Eq.\ref{eq:master}).\\ 
3. The {\em boosted} Born cross section contributes to the total cross 
section in conformity with SF weights in the {\em master} formula 
(Eq.\ref{eq:master}).\\
4. The vacuum polarization effects inserted into all photon propagators 
exactly.\\  
5. Non-leading contributions of order $\alpha$ are accounted for by 
means of the so-called {\em K-factor}.
   
The integration limits of the first term in Eq.\ref{eq:master} were divided 
in two parts from 0 to $\Delta\varepsilon$ 
and from $\Delta\varepsilon$ to the maximal jet energy. 
As a result, the four-fold integral splits into sixteen parts.
Those of them with one photon jet radiation are merged in a proper way 
with {\em compensators} in the {\em master} formula. In this case, 
the total cross section is subdivided into seventeen   
cross sections with own specific kinematics - number of photon jets 
and directions of their radiation. 
The first contribution accounting for effects due to soft and 
virtual radiative corrections is given by
\begin{eqnarray}
&& \frac{\dd\sigma^{e^{+}e^{-} \to e^{+}e^{-} + n\gamma}_{1}}
{\dd\Omega_{1}} = 
\int\limits_{0}^{\Delta}\int\limits_{0}^{\Delta}
\int\limits_{0}^{\Delta}\int\limits_{0}^{\Delta}
\dd x_{1}\dd x_{2}\dd x_{3}\dd x_{4}{\cal D}(z_{1},s) 
{\cal D}(z_{2},s) {\cal D}(z_{3},\tilde{s}){\cal D}(z_{4},\tilde{s}) 
\nonumber \\ && \quad \times \bigl(1 + \frac{\alpha}{\pi}\tilde{K}_{SV}\bigr)
\frac{\dd\tilde{\sigma}_{0}^{e^+e^- \to e^+e^-}(z_{1},z_{2})}{\dd\Omega_{1}} - 
\frac{4\alpha}{\pi}\ln\biggl(\frac{u}{t}\biggr) 
\ln\Delta\frac{\dd\tilde{\sigma}_{0}^{e^+e^- \to e^+e^-}(1,1)}{\dd\Omega_{1}} .
\end{eqnarray}
The photon jet energy emitted by each charged particle can be  
up to $\Delta\varepsilon$. 
This part also contains the contribution due to  production of virtual  
and soft real $e^{+}e^{-}$ pairs if $2m_{e} < \Delta\varepsilon$. 

The next four terms represent the contribution to the cross section with 
one hard jet emission along the motion of any charged particle, 
supplied with the virtual and soft leading logarithmic corrections.  
One of these terms with the relevant {\em compensator} is:
\begin{eqnarray}
&& \frac{\dd\sigma^{e^{+}e^{-} \to e^{+}e^{-} + n\gamma}_{2}}
{\dd\Omega_{1}} = 
\int\limits_{\Delta}^{1}
\int\limits_{0}^{\Delta}
\int\limits_{0}^{\Delta}
\int\limits_{0}^{\Delta}
\dd x_{1}\dd x_{2}\dd x_{3}\dd x_{4} 
{\cal D}(z_{2},s) {\cal D}(z_{3},\tilde{s}){\cal D}(z_{4},\tilde{s})
\frac{\dd\tilde{\sigma}_{0}^{e^+e^- \to e^+e^-}(z_{1},z_{2})}{\dd\Omega_{1}}
\nonumber \\ && 
\biggl[{\cal D}(z_{1},s)\bigl(1+\frac{\alpha}{\pi}\tilde{K}_{SV}\bigr) 
+ \frac{\alpha}{\pi}\frac{1}{x_{1}}
\biggl(\bigl(z_{1} + \frac{x_{1}^2}{2}\bigr)\ln\frac{\theta_{0}^{2}}{4}
 + \frac{x_{1}^2}{2}\biggr)\biggr]\Theta({\mathrm{cuts}}).
\end{eqnarray}
The other similar terms can be written out in the same way by 
the permutation of limits between integrals.   

The next six terms represent the contribution to the cross section when   
two jets are emitted simultaneously along momenta of any two charged 
particles. One of these terms reads
\begin{eqnarray}
&& \frac{\dd\sigma^{e^{+}e^{-} \to e^{+}e^{-} +n\gamma}_{6}}{\dd\Omega_{1}} = 
\int\limits_{\Delta}^{1}
\int\limits_{\Delta}^{1}
\int\limits_{0}^{\Delta}
\int\limits_{0}^{\Delta}
\dd x_{1}\dd x_{2}\dd x_{3}\dd x_{4} 
{\cal D}(z_{1},s) {\cal D}(z_{2},s) {\cal D}(z_{3},\tilde{s}) 
{\cal D}(z_{4},\tilde{s}) 
\nonumber \\ && \quad \times
\frac{\dd\tilde{\sigma}_{0}^{e^+e^- \to e^+e^-}(z_{1},z_{2})}{\dd\Omega_{1}}
\biggl(1 + \frac{\alpha}{\pi}\tilde{K}_{SV}\biggr) \Theta({\mathrm{cuts}}).
\end{eqnarray}
The other similar terms have the identical structure and are obtained by 
the limits permutation.  

The following four terms represent the contribution to the cross 
section when three photon jets are emitted along momenta 
of any three charged particles. One of these terms is given by 
\begin{eqnarray}
&& \frac{\dd\sigma^{e^{+}e^{-} \to e^{+}e^{-} +n\gamma}_{12}}
{\dd\Omega_{1}} = 
\int\limits_{\Delta}^{1}
\int\limits_{\Delta}^{1}
\int\limits_{\Delta}^{1}
\int\limits_{0}^{\Delta}
\dd x_{1}\dd x_{2}\dd x_{3}\dd x_{4} 
{\cal D}(z_{1},s) {\cal D}(z_{2},s) {\cal D}(z_{3},\tilde{s}) 
{\cal D}(z_{4},\tilde{s}) 
\nonumber \\ && \quad \times
\frac{\dd\tilde{\sigma}_{0}^{e^+e^- \to e^+e^-}(z_{1},z_{2})}{\dd\Omega_{1}}
\biggl(1 + \frac{\alpha}{\pi}\tilde{K}_{SV}\biggr) 
\Theta({\mathrm{cuts}}).
\end{eqnarray}

The cross section with emission of four jets along the 
momenta of each initial and final particle is written below,  
\begin{eqnarray}
&& \frac{\dd\sigma^{e^{+}e^{-} \to e^{+}e^{-} +n\gamma}_{16}}
{\dd\Omega_{1}} = 
\int\limits_{\Delta}^{1}
\int\limits_{\Delta}^{1}
\int\limits_{\Delta}^{1}
\int\limits_{\Delta}^{1}
\dd x_{1}\dd x_{2}\dd x_{3}\dd x_{4} 
{\cal D}(z_{1},s) {\cal D}(z_{2},s) {\cal D}(z_{3},\tilde{s}) 
{\cal D}(z_{4},\tilde{s}) 
\nonumber \\ && \quad \times
\frac{\dd\tilde{\sigma}_{0}^{e^+e^- \to e^+e^-}(z_{1},z_{2})}{\dd\Omega_{1}}
\biggl(1 + \frac{\alpha}{\pi}\tilde{K}_{SV}\biggr)\Theta({\mathrm{cuts}}).
\end{eqnarray}

The cross section with one hard photon emission outside the collinear 
region reads
\begin{eqnarray}
\frac{\dd\sigma^{e^{+}e^{-} \to e^{+}e^{-}\gamma}_{17}}{\dd\Omega_{1}} = 
\frac{\alpha^3}{2\pi^2 s}\int\limits_{k^{0}>\Delta\varepsilon 
\atop \theta_{\gamma}>\theta_{0}}R^{e^+e^- \to e^+e^- + \gamma}_{hard}
\frac{s x_{1}x\dd x 
\dd\Omega_{\gamma}}{8(1 - x\sin^2\psi/2)}\Theta({\mathrm{cuts}}),
\end{eqnarray}
where $\psi$ is an angle between the momenta directions of photon 
and final electron.
  
The cross section calculation was performed by the Monte-Carlo method.
The cutoff energy $\Delta\varepsilon$ was chosen at  
ten electron masses to optimise the efficiency of event simulation 
($\Delta = \Delta\varepsilon/\varepsilon \sim 1\%$). Since the {\em master}  
formula is singular on some variables, the main 
of them have been isolated: photon energy and emission 
angle were generated according to functions 
$1/\omega(\varepsilon - \omega)$ and
  $1/(1-\beta_e^2\cos^2\theta_{\gamma})$, respectively. 
The main contribution to the Bhabha cross section comes from the $t$ channel
and it was generated by the function $1/(1-\cos\theta_1)^2$.

The selection criteria adopted here for the simulated events are similar 
to those used in CMD-2 data analysis~\cite{cmd2} and they are:\\
the acollinearity cut in the scattering plane is 
$|\Delta\theta|<0.25$ rad, where $\Delta\theta=\theta_{1}+\theta_{2}-\pi$;\\
the same for azimuthal plane, 
$|\Delta\phi|<0.15$ rad, where $\Delta\phi=|\phi_{1}-\phi_{2}|-\pi$;\\
the angular acceptance is 
$1.1 < \theta_{\mathrm{aver}}< \pi-1.1$, where $\theta_{\mathrm{aver}} 
= (\theta_{1} - \theta_{2} +\pi)/2$; $p^{\bot}_{1,2}>90$~MeV/c.  
Below, if nothing specially told about selection criteria the latter 
will be taken in mind.\\   

The MCGPJ program consists of two main stages. In the first stage 
with a soft selection criteria all majorants for seventeen parts 
are determined, in the second one the cross sections with 
experimental selection criteria are being determined. The generator
simulates an event according to specific kinematics for each cross 
section. The weight of event is determined as the ratio of a given cross 
section (one from seventeen) to the total one. The kinematic parameters 
of the simulated events are stored in the proper histograms, 
which can be compared with experimental distributions.

\begin{figure}[htbp]
  \begin{minipage}[t]{0.45\textwidth}
    \includegraphics[width=0.9701\linewidth]{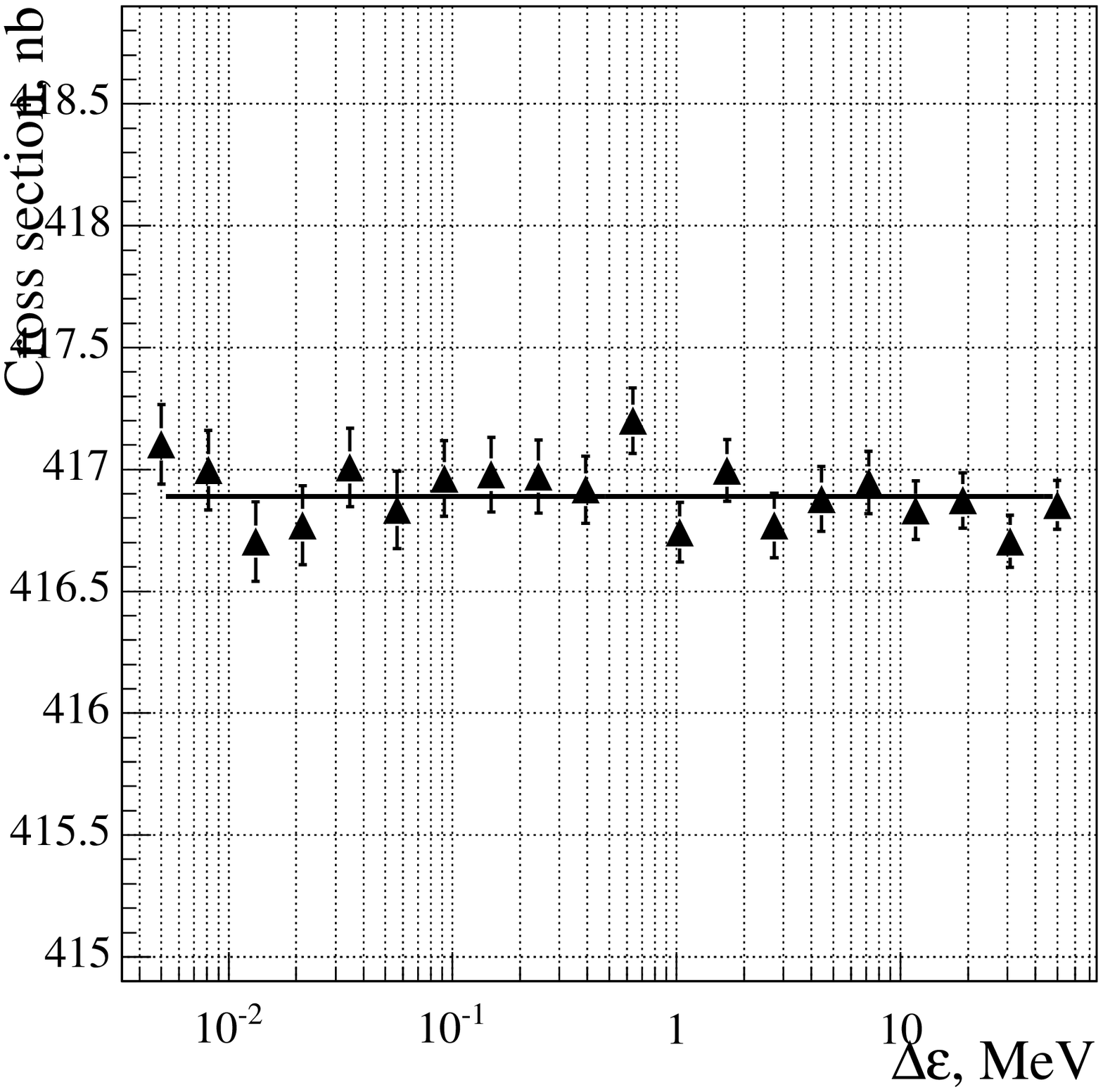}
    \caption{\label{fig:de} The cross section dependence on  
the auxiliary parameter $\Delta\varepsilon$. Parameters and selection 
cuts are given in text.}
  \end{minipage}
  \hfill
  \begin{minipage}[t]{0.45\textwidth}
    \includegraphics[width=0.9701\linewidth]{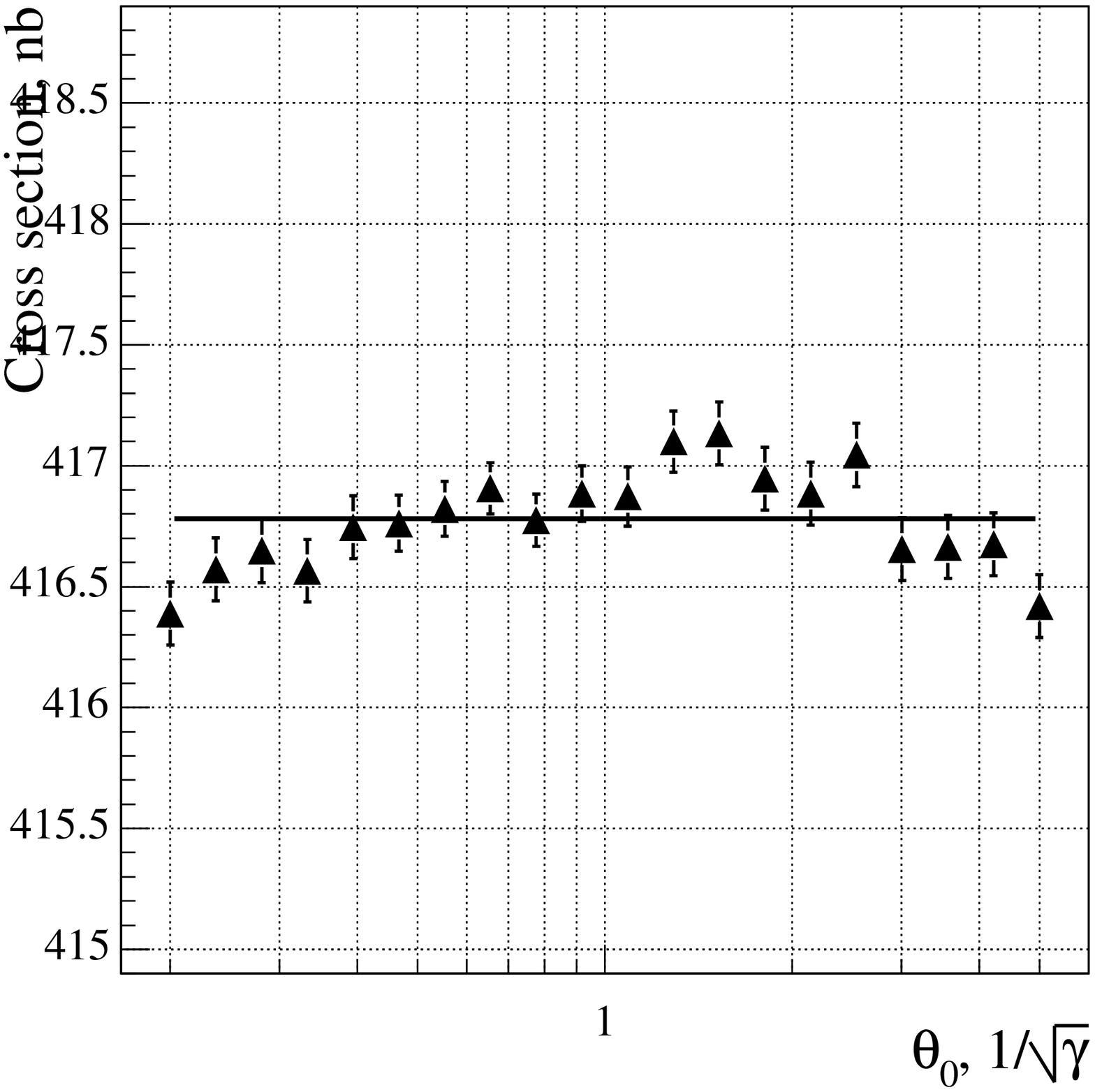}
    \caption{\label{fig:th0} The cross section dependence on  
the auxiliary parameter $\theta_{0}$ after integration over the 
remaining kinematic variables.} 
  \end{minipage}
\end{figure}

Numerous tests have been performed for the c.m.s. energy 
of 900~MeV. The cross section dependence on  
the auxiliary parameter $\Delta\varepsilon$ is shown in 
Fig.\ref{fig:de} after integration over the remaining kinematic 
variables. It is seen that the cross section variations are inside the 
claimed precision while $\Delta\varepsilon$ changes by a factor of 10$^4$.
The cross section variations with an auxiliary parameter  
$\theta_{0}$ do not exceed $\pm 0.1\%$ level (Fig.\ref{fig:th0}). From that 
certainly follows that the cross section stability to the auxiliary parameters 
$\Delta\varepsilon$ and $\theta_{0}$ at least is not worse than 0.1\%.  

\begin{figure}[htbp]
  \begin{minipage}[t]{0.45\textwidth}
    \includegraphics[width=0.9701\linewidth]{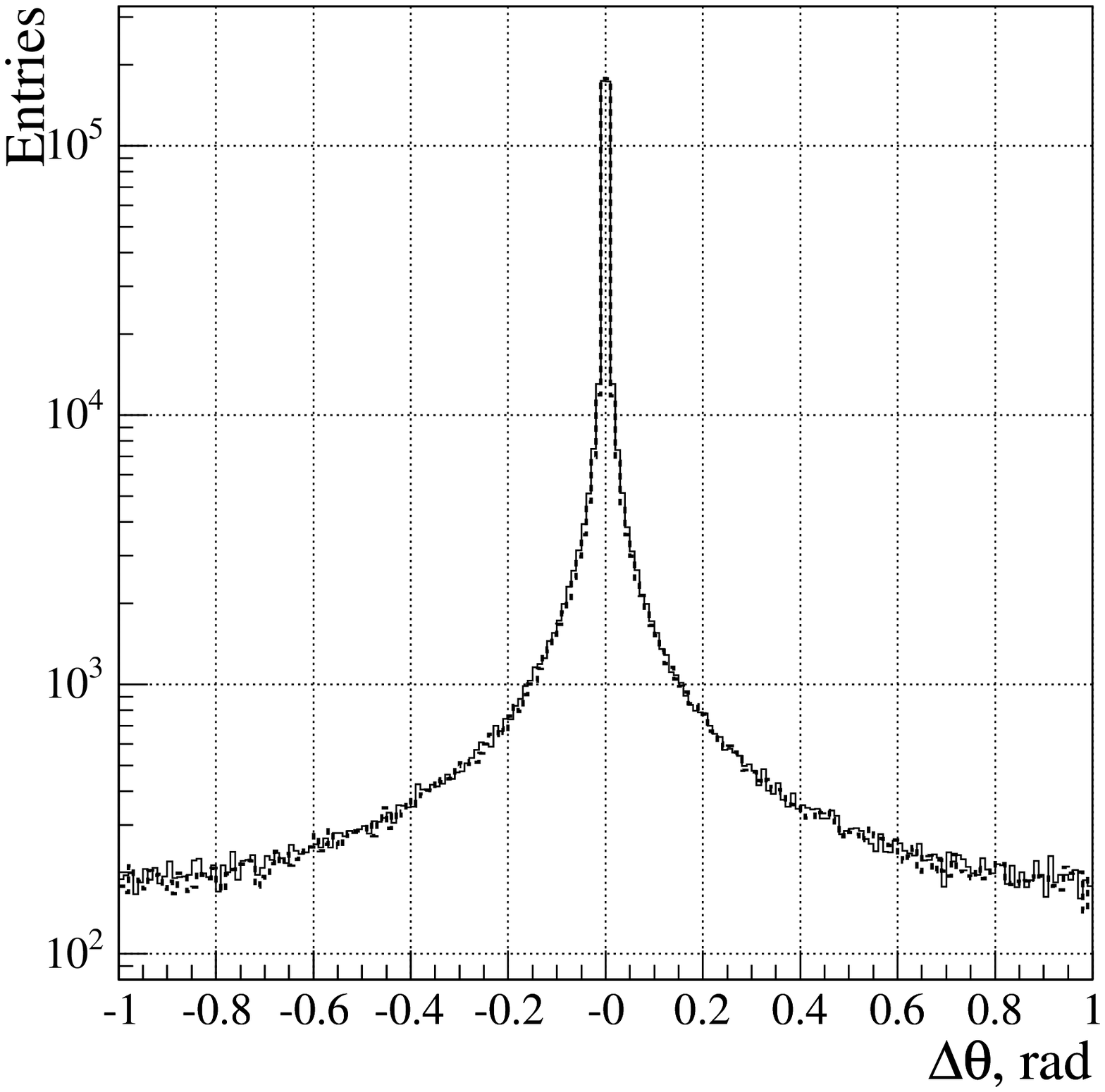}
    \caption{\label{fig:dtheta} The acollinearity 
polar angle $\Delta\theta$ distribution. The solid line--\mbox{MCGPJ} code, 
the dashed line--\mbox{BHWIDE}.}
  \end{minipage}
  \hfill
  \begin{minipage}[t]{0.45\textwidth}
    \includegraphics[width=0.9701\linewidth]{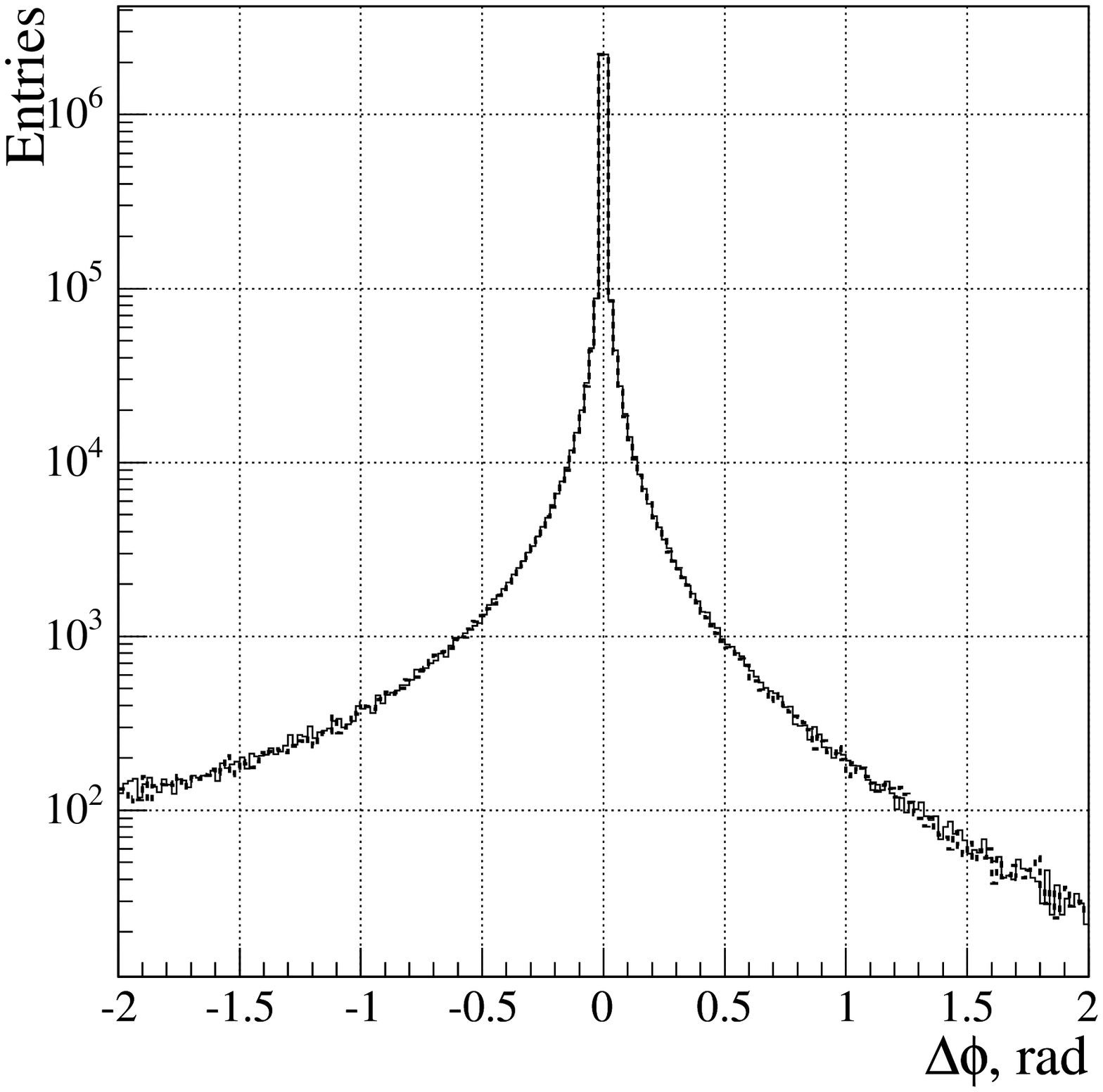}
    \caption{\label{fig:dphi} The acollinearity
azimuthal angle $\Delta\phi$ distribution. The solid line -- 
\mbox{MCGPJ} code, the dashed line -- \mbox{BHWIDE}.}
  \end{minipage}
\end{figure}

Comparison of different kinematic distributions simulated by the 
\mbox{MCGPJ} generator and \mbox{BHWIDE}~\cite{bhwide} was performed. 
Distributions over parameters $\Delta\theta$ = $\theta_{1}+
\theta_{2}-\pi$ and $\Delta\phi$ = $|\phi_{1}-\phi_{2}|-\pi$ 
are plotted in Figs.\ref{fig:dtheta},~\ref{fig:dphi}.
Good agreement  
is seen while $\Delta\theta$ and $\Delta\phi$ vary in the wide range. 
Asymmetric shape of the distribution on $\Delta\phi$ is due to 
specific kinematic of the events, when photon and 
$e^+e^-$ pair fly in diametrically opposed sides. The cross section 
for such kinematic has enhancement which visualizes as shape asymmetry   
of the distribution.   

\begin{figure}[htbp]
  \begin{minipage}[t]{0.45\textwidth}
\includegraphics[width=0.9501\textwidth]{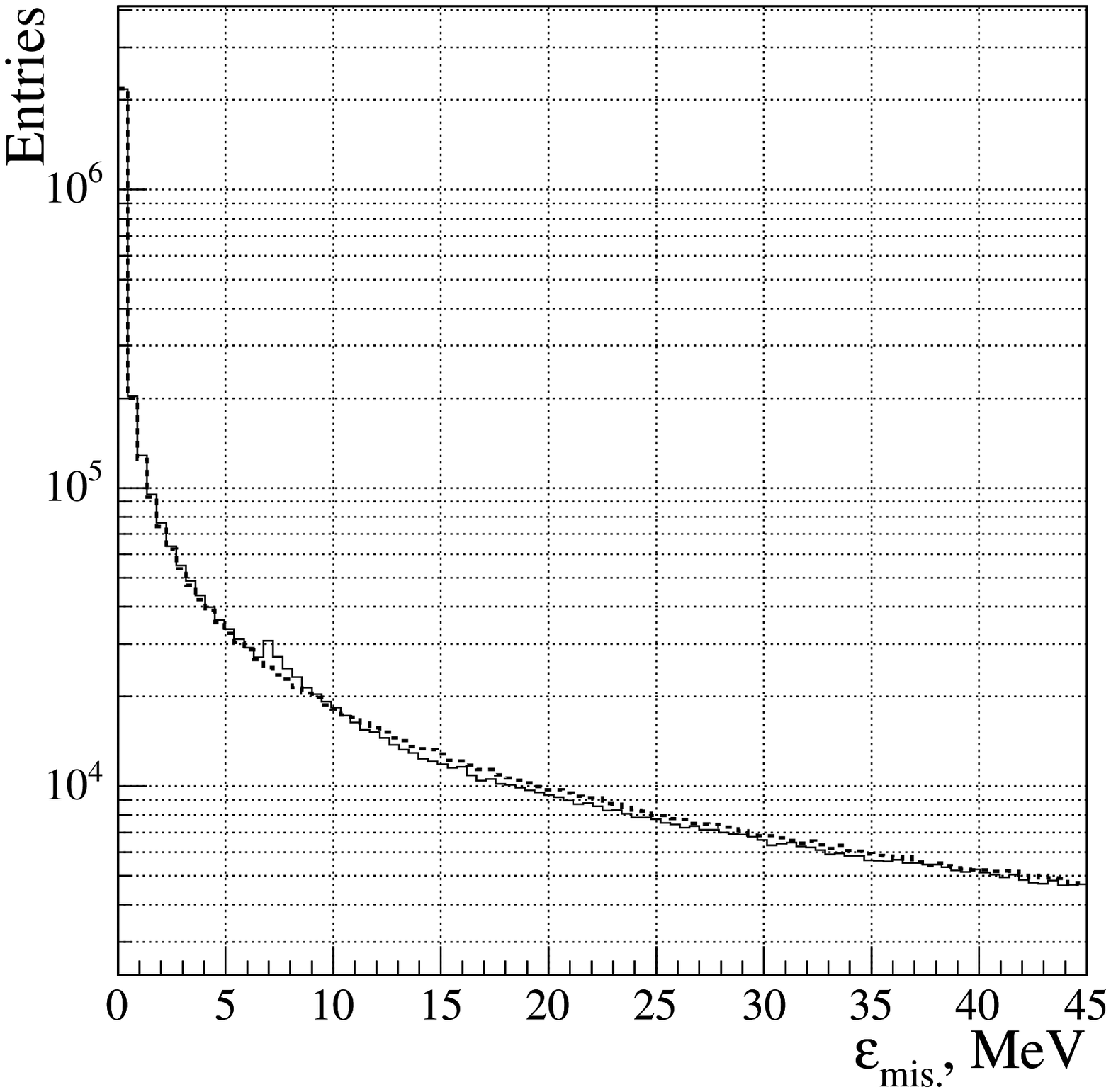}
\caption{\label{fig:phen} The distribution of events as a function of the  
missing energy radiated by electrons and positrons. 
The solid line -- \mbox{MCGPJ} code, the dashed line -- \mbox{BHWIDE}.}
  \end{minipage}
  \hfill
  \begin{minipage}[t]{0.45\textwidth}
  \includegraphics[width=0.9501\textwidth]{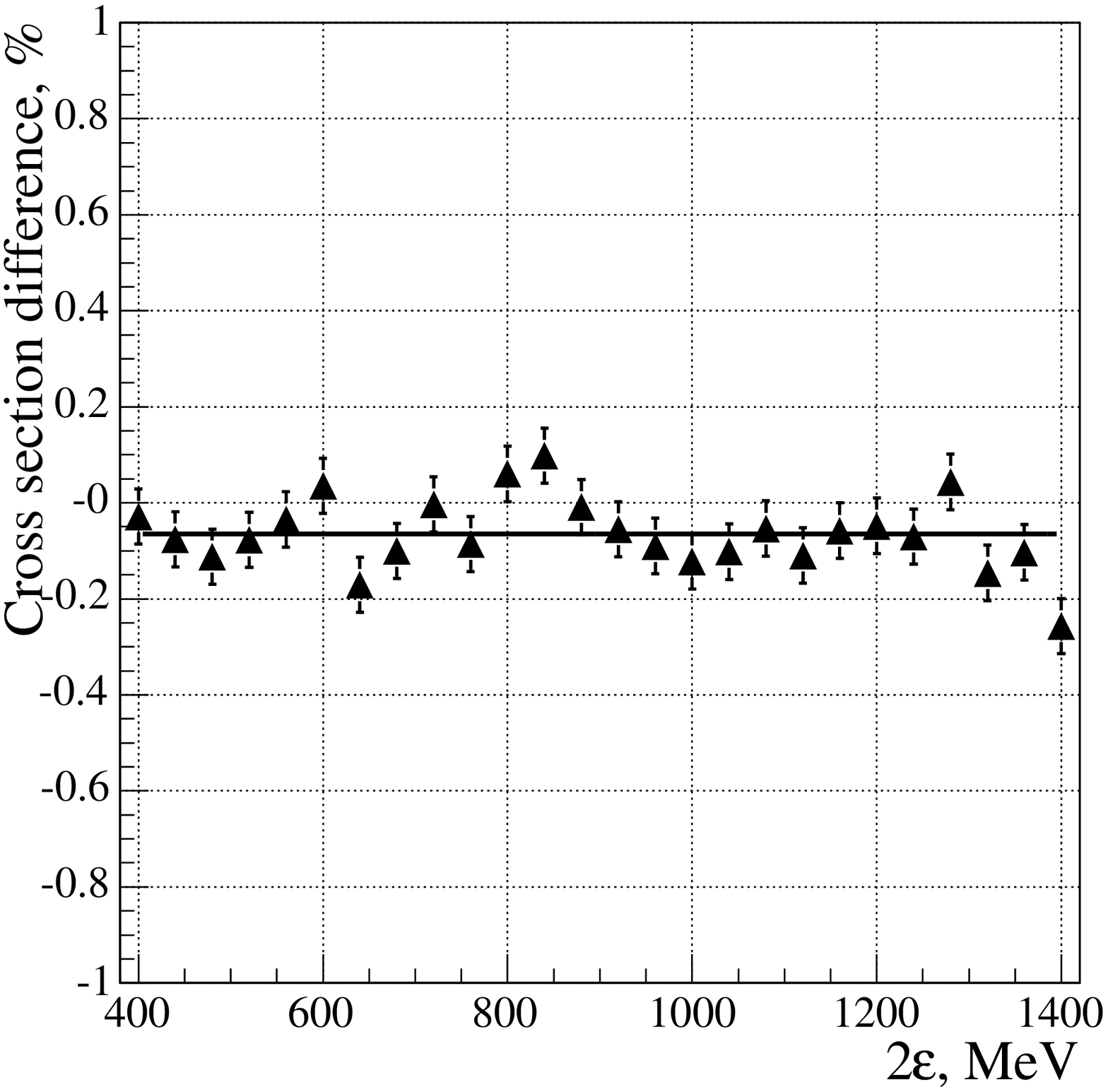}
  \caption{\label{fig:escan} The relative difference of cross sections 
calculated by the \mbox{MCGPJ} code and \mbox{BHWIDE} as a function of 
the c.m.energy.}
 \end{minipage}
\end{figure}

The event distributions produced by both generators are presented 
in Fig.~\ref{fig:phen} as a function of missing energy 
($\varepsilon_{mis.} =2\varepsilon - \varepsilon_{1} - \varepsilon_{2} $). 
Both distributions are close to each other except for the energy region where 
soft and hard photons are merged. A visible bump is observed in this point. 
This bump originates from a slightly different dependence 
on the cutoff energy of both the {\em compensators} and the cross section 
with one hard photon. The cut on acollinearity  
$\Delta\theta \sim$ 0.25~rad, is equivalent to missing energy  
$\sim$ 100~MeV that is rather far from cutoff energy $\Delta\varepsilon 
\sim $ 7~MeV. As a result, the contribution into the total cross section 
connected with this spurious features is negligible.  

\begin{figure}[htbp]
  \begin{minipage}[t]{0.45\textwidth}
    \includegraphics[width=0.9701\textwidth]{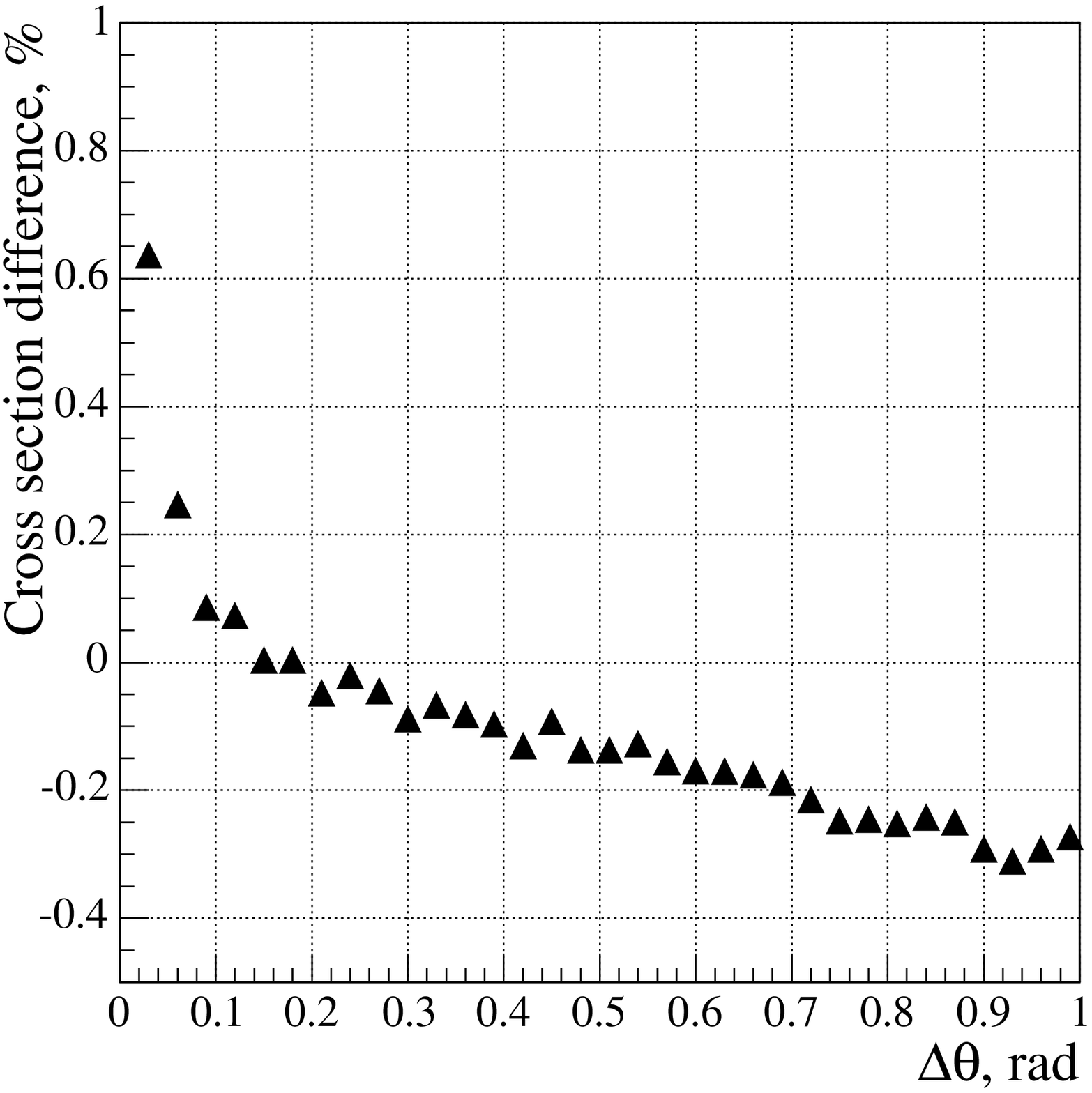}
    \caption{\label{fig:diffvsdth} The relative difference between cross
sections calculated by the \mbox{MCGPJ} code and \mbox{BHWIDE} {\it versus}
      the acollinearity angle $|\Delta \theta |$.}
  \end{minipage}
  \hfill
  \begin{minipage}[t]{0.45\textwidth}
    \includegraphics[width=0.9701\textwidth]{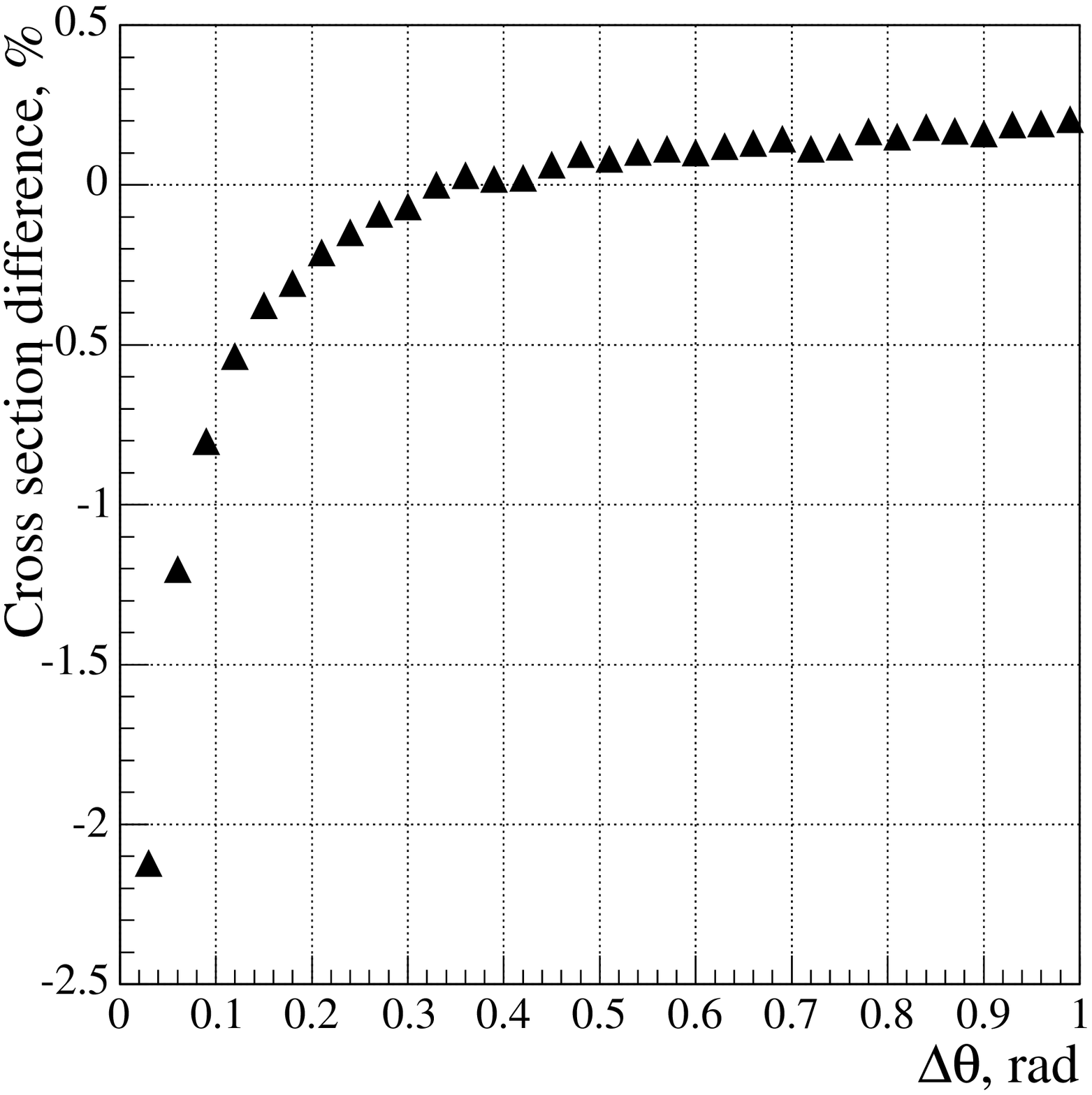}
    \caption{\label{fig:diffvsdth-one-ph} The relative difference
      between cross sections calculated by the \mbox{MCGPJ} code and the
      generator based on Ref.~\cite{berends} {\it versus} 
the acollinearity angle $|\Delta \theta |$.}
  \end{minipage}
\end{figure}

The relative difference of the cross sections, presented 
in Fig.\ref{fig:escan}, calculated by the \mbox{MCGPJ} code and 
\mbox{BHWIDE} is less than 0.1\% for the VEPP-2M energy range. 
This difference {\em versus} the acollinearity 
angle $\Delta \theta$ is plotted in Fig.\ref{fig:diffvsdth}. One can see 
that the size and sign of the difference depend on the particular choice of 
$\Delta \theta$. The reason of the difference about $\sim 0.5\%$ for 
$\Delta\theta \sim 0.05$~rad arises from the fact that all photons 
(except one in our code) are emitted strongly along the motion of electrons 
(positrons) whereas in \mbox{BHWIDE} they have some angular distribution.
The difference of $\sim 0.3\%$ for the large acollinearity angles 
$|\Delta \theta| 
\sim$ 1~rad is due to the fact that the \mbox{BHWIDE} code simulates one 
hard photon only. It is worth noticing the \mbox{MCGPJ} code describes  
the shape of {\em tails} of the different kinematic distributions  
a bit correctly and, as a result, it is preferable for applications 
when {\em soft} selection criteria are used. 

It is important to reliably estimate the total theoretical precision 
of this approach. In order to quantify a theoretical error,
the independent comparison has been performed with the generator based
on Ref.~\cite{berends}, where $\mathcal O(\alpha)$ QED corrections are
treated exactly. It was found that the relative difference of cross sections
is more than 1\% for small acollinearity angles $\Delta\theta<0.1$~rad
( Fig.~\ref{fig:diffvsdth-one-ph} ) and it is less than $\sim 0.2\%$ 
for acollinearity angles $\sim$ 0.25~rad.  From that immediately follows 
that the radiation of two and more photons (jets) in 
the collinear region contributes to the cross section by the amount 
$\sim 0.2\%$ only. Therefore, we can conclude that the theoretical 
precision of the Bhabha cross section with RC is certainly better 
than $\sim 0.2\%$ for our selection criteria.

\begin{figure}[htbp]
  \begin{minipage}[t]{0.45\textwidth}
    \includegraphics[width=0.9501\textwidth]{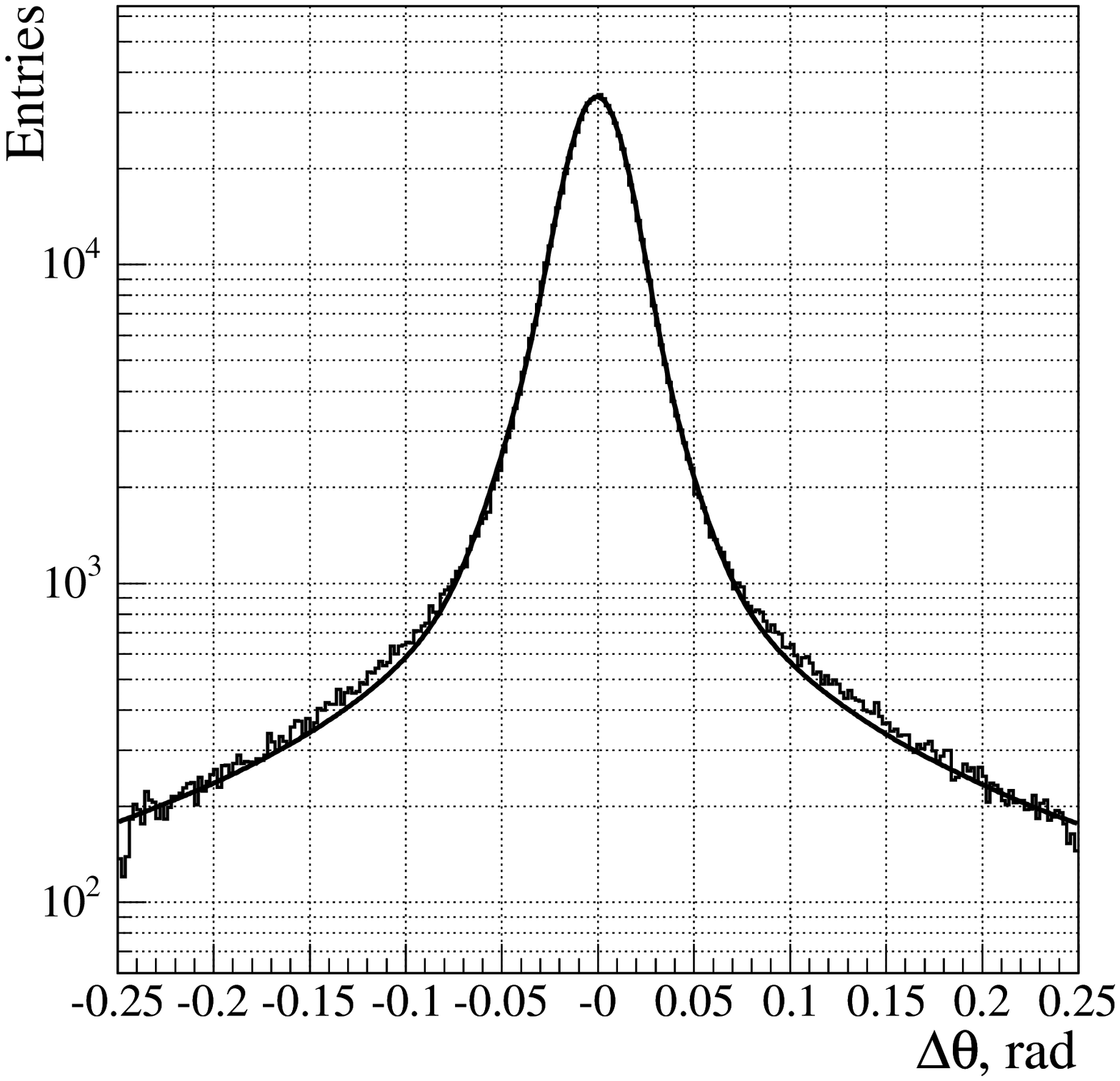}
    \caption{\label{fig:sim-exp-dth} The acollinearity angle $\Delta\theta$ 
distribution in the scattering
      plane. Solid line - simulation, histogram - experiment.}
  \end{minipage}
  \hfill
  \begin{minipage}[t]{0.45\textwidth}
    \includegraphics[width=0.9501\textwidth]{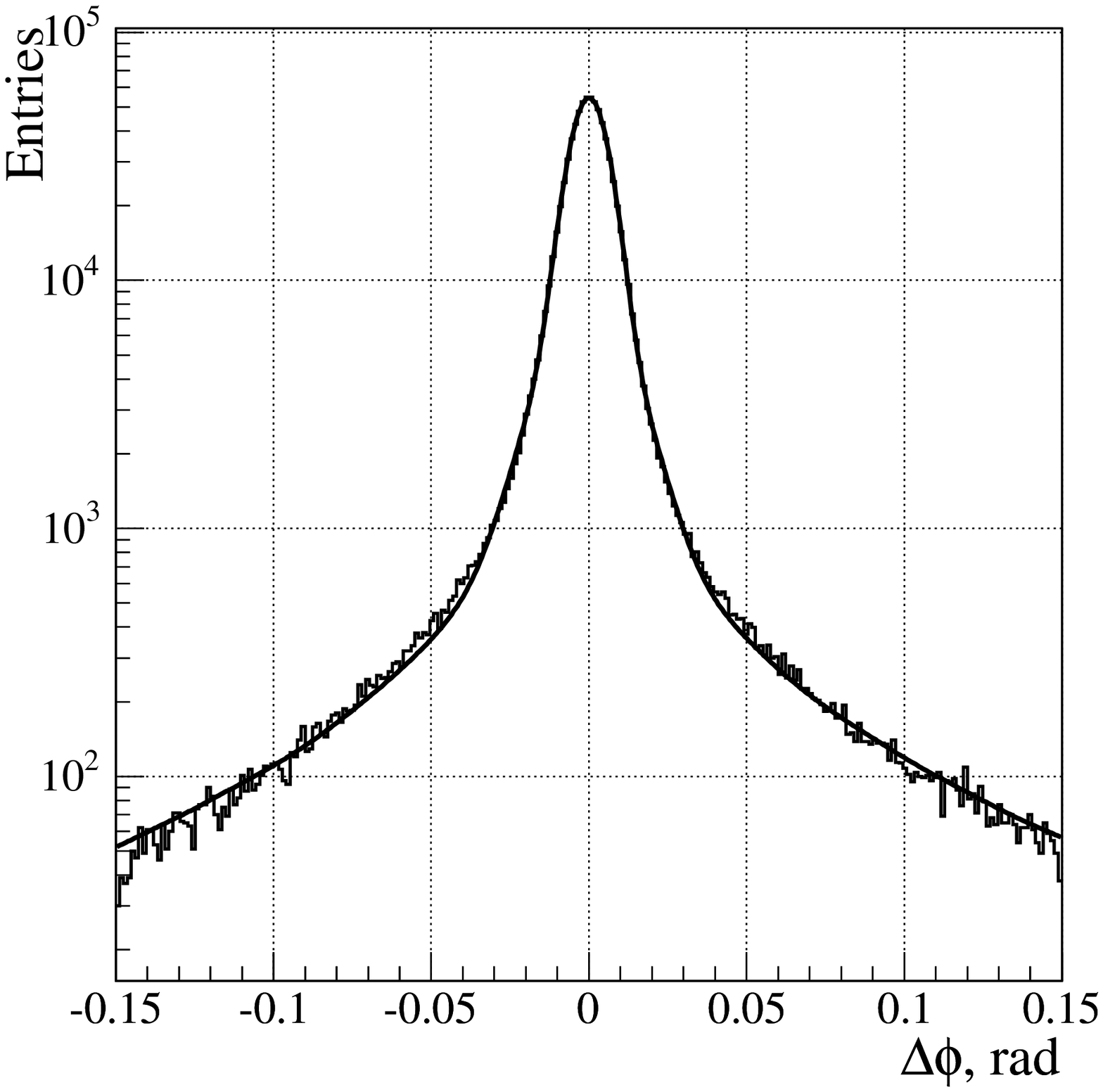}
    \caption{\label{fig:sim-exp-dphi} The acollinearity angle $\Delta\phi$ 
distribution in the azimuthal plane. Solid
      line - simulation, histogram - experiment.}
  \end{minipage}
\end{figure}

The EM calorimeter of the CMD-2 detector allows to separate Bhabha scattering
events from others with high confidence level~\cite{cmd2}. The distributions 
in acollinearity angles $\Delta\theta$ and $\Delta\phi$ are presented in 
Figs.\ref{fig:sim-exp-dth},\ref{fig:sim-exp-dphi}. 
To increase the experimental statistics, all the CMD-2 
data at energies greater than 1040~MeV collected in these
plots. The momentum and angular
resolutions, interaction with the detector material
were added to the kinematic parameters of simulated events. 
The histograms were fitted by two Gaussian functions. 
Their relative weights and widths were fit parameters. 
Good agreement between experiment and simulation can be seen.

\begin{figure}[htbp]
\begin{minipage}[t]{0.45\textwidth}
  \includegraphics[width=0.9501\textwidth]{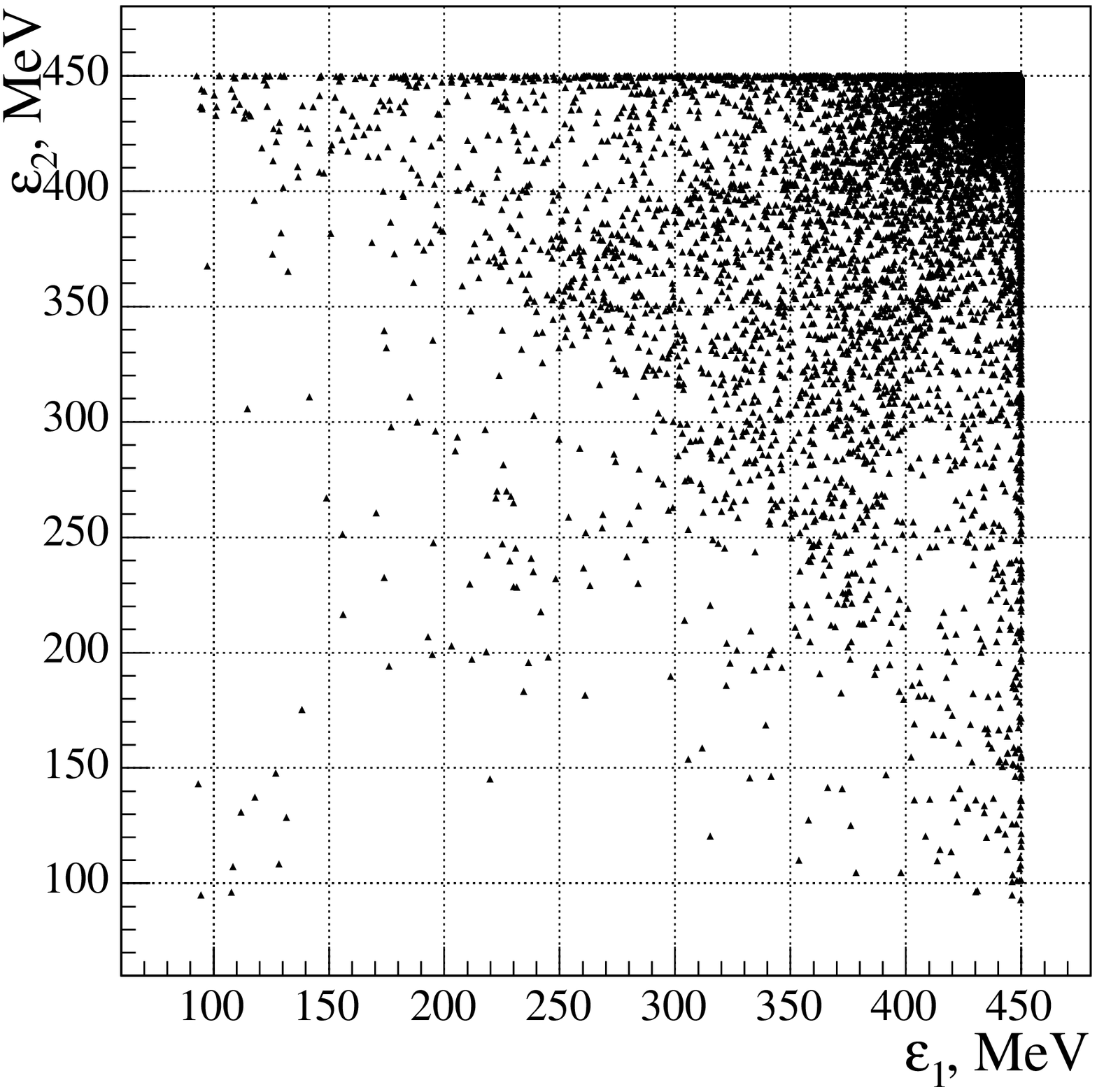}
  \caption{\label{fig:e1e2a} Two-dimensional plot of the simulated 
events (MCGPJ). The points in this plot correspond to 
the electron and positron energies. The influence of the condition 
$\Delta\theta < 0.25$~rad can be seen from an ark-like smooth border.}
\end{minipage}
\hfill
\begin{minipage}[t]{0.45\textwidth}
  \includegraphics[width=0.9501\textwidth]{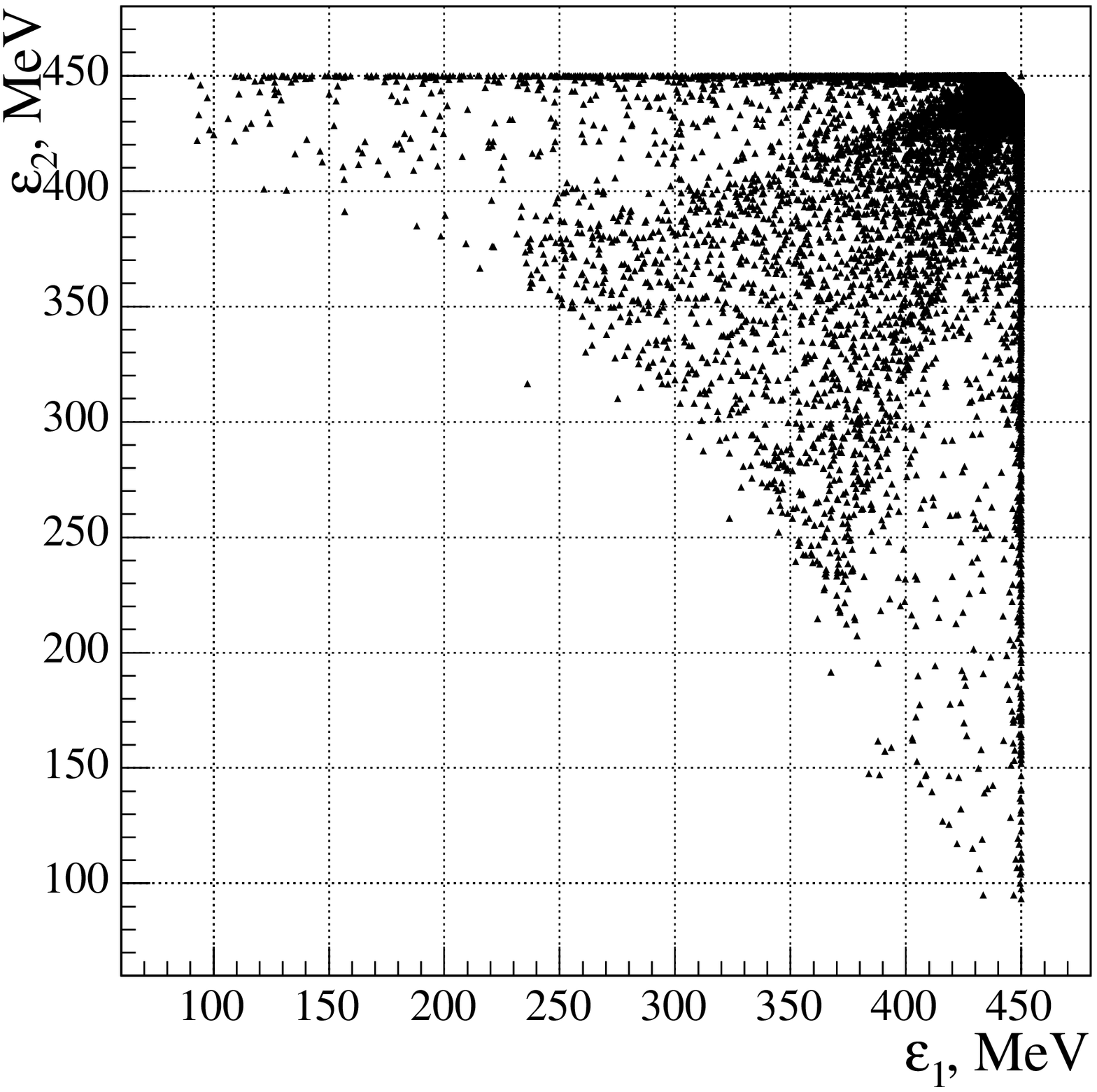}
  \caption{\label{fig:e1e2b}Two-dimensional plot of the simulated 
events. The generator is based on~\cite{berends}. 
The points in this plot correspond to the electron and positron 
energies. The condition $\Delta\theta < 0.25~rad$ 
divides the plot into two parts - with and without events.}
\end{minipage}
\end{figure}

\begin{figure}[htbp]
\begin{minipage}[t]{0.45\textwidth}
  \includegraphics[width=0.9501\textwidth]{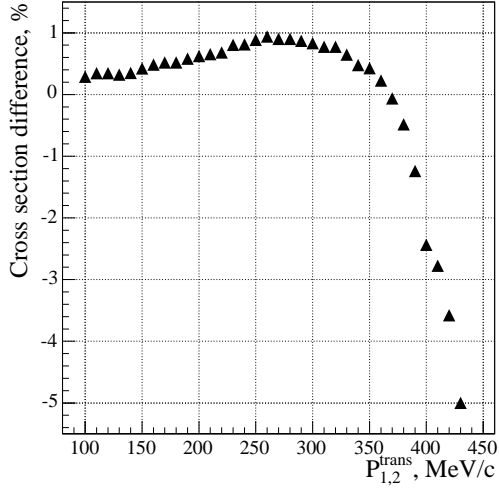}
  \caption{\label{fig:csd-pcut} The difference between cross sections 
as a function of the cut imposed on the  transverse momenta of final 
particles.}
\end{minipage}
\end{figure}

The agreement between experiment and simulation becomes significantly 
worse when the MC generator based on Ref.~\cite{berends} with 
${\mathcal O}(\alpha)$ 
corrections is used. It is seen in Figs.\ref{fig:e1e2a},\ref{fig:e1e2b}
where two-dimensional plots are presented. The points in these plots 
correspond to the electron and positron energies.  
Different population of events is observed far aside from the area where  
{\em semi-elastic} events are concentrated. About $\sim 1\% $ events 
have correlated low energies and they are distributed predominantly 
along a corridor which extends from the right upper to the left bottom 
corner of this plot. The appearance of these events is  
due to simultaneous radiation of two jets with close 
energies along either initial or final particles. The condition  
$p^{\bot}_{1,2} > $90~MeV/c is very soft and only owing to this fact the 
integrated cross sections are equal to each other within $\sim 0.2\%$. 
If $p^{\bot}_{1,2}$ is about $\sim 220$~MeV/c, the relative difference 
increases up to $\sim 1\%$ (Fig.\ref{fig:csd-pcut}).     
For the values $p^{\bot}_{1,2}$ more than 350~MeV/c the difference changes 
a sign and grows up. The cross section with photon jets becomes 
smaller than with one photon under condition $p^{\bot}_{1,2} > $
350~MeV/c. 
This feature has a simple explanation. The distribution width of 
{\em semi-elastic} events in the first plot is broader than for the second 
one due to radiation of many soft photons and, as a result, these  
events are {\em smeared} more broadly near the peak area.   

\section{Monte-Carlo generator for production of muon pairs }
The same approach was used to create the MC generator 
to simulate production of muon pairs in the reaction  
$e^{-}(z_{1}p_{-}) + 
e^{+}(z_{2}p_{+}) \to \mu^{-}(p_{1}) + \mu^{+}(p_{2}),$ 
when initial particles radiate some energy by emission of photon jets  
in the collinear region. According to~\cite{arkurlept} the {\em boosted} 
Born cross section modified by the vacuum polarization effects in the 
photon propagator reads
\begin{eqnarray}
\frac{\dd\tilde\sigma^{e^+e^- \to \mu^+\mu^-}_{0}(z_{1},z_{2})}
{\dd\Omega_{1}} = 
\frac{\alpha^2}{4s}
\frac{1}{\mid 1 - \Pi(z_{1}z_{2}s) \mid ^2}
\frac{y_{1}\bigl[z_{1}^2(Y_{1} - y_{1}c_{1})^2 + z_{2}^2(Y_{1} + 
y_{1}c_{1})^2 + 8z_{1}z_{2}m_{\mu}^2/s\bigr]}
{z_{1}^3z_{2}^3\bigl[z_{1} + z_{2} - (z_{1} - z_{2})c_{1}Y_{1}/y_{1}\bigr]},
\end{eqnarray}       
where $y_{1,2}^2 = Y_{1,2}^2 - 4m^2_{\mu}/s$;
$Y_{1,2} = \varepsilon_{1,2}/\varepsilon$ are the muon relative energies;
$z_{1,2} = 1 - x_{1,2},$ 
$ x_{1,2} = \omega_{1,2}/\varepsilon$ are the relative energies of photon jets;
$c_{1} = \cos\theta_{1}$, $\ \theta_{1}$ is a polar angle of negative muon.  
The energy-momentum conservation law, 
\begin{center}
$z_{1}+z_{2}=Y_{1}+Y_{2}$, 
$\hspace{0.2 cm} z_{1}-z_{2}=y_{1}c_{1}+y_{2}c_{2},
\hspace{0.2 cm} y_{1}\sqrt{1-c_{1}^2}=y_{2}\sqrt{1-c_{2}^2},$  
\end{center}
allows to determine $Y_{1}$, $Y_{2}$ and a positron polar angle 
$\theta_{2}$ $(c_{2} = \cos{\theta_{2}})$: 
\begin{eqnarray}
Y_1=\frac{2m_{\mu}^2}{s}
\frac{(z_2-z_1)c_{1}}{z_1z_2+[z_1^2z_2^2-(m_{\mu}^2/s)((z_1+z_2)^2
-(z_1-z_2)^2c_{1}^2)]^{1/2}}
+\frac{2z_1z_2}{z_1+z_2-c_{1}(z_1-z_2)}.
\end{eqnarray}
 
The charge-even part of the cross section in the first order in $\alpha$
comes from one-loop virtual~(V) and soft~(S) 
radiative corrections and according to Ref.~\cite{csmumu} 
is given by: 
\begin{eqnarray}
\frac{\dd\sigma^{S+V}_{even}}{\dd\Omega_{1}} = 
\frac{\dd\tilde\sigma^{e^+e^- \to \mu^+\mu^-}_{0}(1,1)}{\dd\Omega_{1}}
\frac{2\alpha}{\pi} (A_{e} + A_{\mu}), 
\nonumber\\ 
A_{e} = (L - 1)\ln\frac{\Delta\varepsilon}{\varepsilon} + 
\frac{3}{4}(L - 1) + \frac{\pi^2}{6} - \frac{1}{4}, \hspace{4 mm} 
A_{\mu} = \biggl(\frac{1 + \beta^2}{2\beta}\ln\frac{1 + \beta}{1 - \beta} 
- 1 \biggr)\ln\frac{\Delta\varepsilon}{\varepsilon} + 
K_{\mathrm{even}}^{\mu}.
\end{eqnarray}
The expression for the quantity $K_{\mathrm{even}}^{\mu}$ was derived 
in~\cite{khrip} and reads
\begin{eqnarray}
K_{\mathrm{even}}^{\mu} &=& -1 
+ \rho\biggl(\frac{1+\beta^2}{2\beta} - \frac{1}{2}
+ \frac{1}{4\beta}\biggr) + \ln\frac{1+\beta}{2}\left(\frac{1}{2\beta}
+ \frac{1+\beta^2}{\beta}\right)  
\nonumber \\
&-& \frac{1-\beta^2}{2\beta}\frac{l_\beta}{2-\beta^2(1-c_{1}^2)}
+ \frac{1+\beta^2}{2\beta}\biggl[ \frac{\pi^2}{6}
+ 2\mbox{Li}_2\left(\frac{1-\beta}{1+\beta}\right)
+ l_{\beta}\ln\frac{1+\beta}{2\beta^2}\biggr], 
\nonumber \\
l_{\beta} &=& \ln\frac{1+\beta}{1-\beta}\, , \quad
\rho=\ln\frac{s}{m_{\mu}^2}\, ,\quad 
L=\ln\frac{s}{m_{e}^2}\, , \quad
\mbox{Li}_2(x)\equiv -\int\limits_{0}^{x}\frac{\dd t}{t}\ln(1-t).
\end{eqnarray}
In the ultra relativistic limit ($\beta \to 1$) the quantity $A_{\mu}$ 
takes the same form as for electrons  
\begin{eqnarray}
A_{\mu} = (L_{\mu} - 1)\ln\frac{\Delta\varepsilon}{\varepsilon} + 
\frac{3}{4}(L_{\mu} - 1) + \frac{\pi^2}{6} - \frac{1}{4}.
\end{eqnarray}

The charge-odd part of the cross section 
comes from the interference of the Born amplitude with 
box-type diagrams as well as with amplitudes of soft photon emission by
initial and final particles and is given by~\cite{khrip,arkurlept}:  
\begin{eqnarray}
\frac{\dd\sigma^{S+V}_{\mathrm{odd}}}{\dd\Omega_{1}} &=& 
\frac{\dd\sigma^{e^+e^- \to \mu^+\mu^-}_{0}(1,1)}{\dd\Omega_{1}}
\frac{2\alpha}{\pi}
\biggl(2\ln\frac{\Delta\varepsilon}{\varepsilon} 
\ln\frac{1 - \beta c_{1}}{1 + \beta c_{1}} + 
K_{\mathrm{odd}}^{\mu}\biggr), 
\nonumber\\
K_{\mathrm{odd}}^{\mu} &=& \frac{1}{2}l_-^2 - L_-(\rho+l_-)
+ \mbox{Li}_2\left(\frac{1-\beta^2}{2(1-\beta c_{1})}\right)
+ \mbox{Li}_2\left(\frac{\beta^2(1-c_{1}^2)}
{1+\beta^2-2\beta c_{1}}\right)
\nonumber \\
&-& \int\limits_{0}^{1-\beta^2}\frac{\dd x}{x}f(x)
\biggl(1-\frac{x(1+\beta^2-2\beta c_{1})}{(1-\beta c_{1})^2}
\biggr)^{-\frac{1}{2}} + \frac{1}{2-\beta^2(1-c_{1}^2)}
\nonumber \\
&\times& \Biggl\{ - \frac{1-2\beta^2+\beta^2 c_{1}^2}
{1+\beta^2-2\beta c_{1}}(\rho+l_-) - \frac{1}{4}(1-\beta^2)
\biggl[l_-^2 - 2L_-(l_-+\rho) 
\nonumber \\
&+& 2\mbox{Li}_2\left(\frac{1-\beta^2}{2(1-\beta c_{1})}\right)\biggr]
+ \beta c_{1}\biggl[-\frac{\rho}{2\beta^2} + \biggl(\frac{\pi^2}{12}
+ \frac{1}{4}\rho^2\biggr)\biggl(1- \frac{1}{\beta} - \frac{\beta}{2}
+ \frac{1}{2\beta^3}\biggr)
\nonumber \\
&+& \frac{1}{\beta}(-1-\frac{\beta^2}{2}
+ \frac{1}{2\beta^2}) \biggl(\rho\ln\frac{1+\beta}{2}
- 2\mbox{Li}_2\left(\frac{1-\beta}{2}\right)
- \mbox{Li}_2\left(-\frac{1-\beta}{1+\beta}\right)\biggr) 
\nonumber \\
&-& \frac{1}{2}l_-^2 + L_-(\rho+l_-)
- \mbox{Li}_2\left(\frac{1-\beta^2}{2(1-\beta c_{1})}\right) \biggr]\Biggr\}
- (c_{1}\rightarrow -c_{1}), \\ 
\nonumber
f(x)&=&\biggl(\frac{1}{\sqrt{1-x}}-1\biggr)
\ln\frac{\sqrt{x}}{2} - \frac{1}{\sqrt{1-x}}\ln\frac{1+\sqrt{1-x}}{2}\, ,\\ 
\nonumber
l_-&=&\ln\frac{1-\beta c_{1}}{2}\, ,\qquad
L_-=\ln\biggl(1-\frac{1-\beta^2}{2(1-\beta c_{1})}\biggr).
\end{eqnarray}
For the ultra relativistic limit the same result as in~\cite{khrip} 
is obtained.

The cross section of muon pair production with one hard photon
emission is studied in detail elsewhere~\cite{csmumu, arkurlept}. 
This cross section in the differential form,
keeping the relevant information about the  kinematics of final 
particles, can be written as:
\begin{eqnarray}
\dd\sigma^{e^+e^- \to \mu^+\mu^-\gamma}_{\mathrm{hard}} = 
\frac{\alpha^3}{2\pi^2s^2}
R^{e^+e^- \to \mu^+\mu^-\gamma}_{\mathrm{hard}} 
\frac{s\beta_{1}\dd\Omega_{1}x\dd x\dd\Omega_{\gamma}}
{4[2 - x(1-\cos\psi/\beta_{1})]}, 
\end{eqnarray} 
where $\beta_{1}$ is a velocity of negative muon.
The quantity $R^{e^+e^- \to \mu^+\mu^-\gamma}_{\mathrm{hard}}$
consists of three terms and represents the cross section with one hard
photon emitted by the initial and final particles as well as their 
interference:
\begin{eqnarray}
&& R^{e^+e^- \to \mu^+\mu^-\gamma}_{\mathrm{hard}}=\frac{s}
{16(4\pi\alpha)^3}\sum\limits_{spins}^{}
|M|^2=R_{ee} + R_{e\mu} + R_{\mu\mu}, 
\\ \nonumber 
&& R_{ee} = \frac{1}{|1 - \Pi(s_{1})|^2}\left[C
\frac{s}{\chi_-\chi_+} + \frac{m_{\mu}^2}{s_1^2}\Delta_{s_1s_1}  
-\frac{m_e^2}{2\chi_-^2}\frac{(t_1^2+u_1^2+2m_{\mu}^2s_1)}{s_1^2} 
 - \frac{m_e^2}{2\chi_+^2}\frac{(t^2+u^2+2m_{\mu}^2s_1)}{s_1^2}\right], \\ 
\nonumber  
&& R_{e\mu} = \Ree\frac{1}{(1 - \Pi(s_{1}))(1 - \Pi(s))^{*}}\left[ 
C( \frac{u}{\chi_-\chi_+'} + \frac{u_1}{\chi_+\chi_-'}
   - \frac{t}{\chi_-\chi_-'} - \frac{t_1}{\chi_+\chi_+'}) 
+\frac{m_{\mu}^2}{ss_1} \Delta_{ss_1}\right] \\ 
\nonumber  
&& R_{\mu\mu} = \frac{1}{|1 - \Pi(s)|^2}\left[
\frac{s_1}{\chi_-'\chi_+'}C + \frac{m_{\mu}^2}{s^2} \Delta_{ss}\right]
\, , \quad C = \frac{u^2+u_1^2+t^2+t_1^2}{4ss_1}\, ,\\ 
\nonumber && 
\Delta_{s_1s_1} = \frac{(t+u)^2+(t_1+u_1)^2}{2\chi_-\chi_+}\, ,\quad 
\Delta_{ss} = - \frac{u^2+t_1^2+2sm_{\mu}^2}{2(\chi_-')^2} 
- \frac{u_1^2+t^2+2sm_{\mu}^2}{2(\chi_+')^2} \\ 
\nonumber
&& + \frac{( ss_1 - s^2 + tu +t_1u_1 - 2sm_{\mu}^2)}{\chi_-'\chi_+'}, 
\\ \nonumber
&& \Delta_{ss_1} = \frac{s+s_1}{2}\biggl( \frac{u}{\chi_-\chi_+'}
+ \frac{u_1}{\chi_+\chi_-'} - \frac{t}{\chi_-\chi_-'}
- \frac{t_1}{\chi_+\chi_+'} \biggr) 
+ \frac{2(u-t_1)}{\chi_-'} + \frac{2(u_1-t)}{\chi_+'}.
\end{eqnarray}
Mandelstam variables and ~$\chi_{\pm},$ $~\chi_{\pm}'$ are defined 
as for electrons. 
Similar to the Bhabha cross section, the {\em master} formula 
describing the process of muon pair production reads~\cite{arkurlept} 
\begin{eqnarray}  
&& \frac{\dd\sigma^{e^+e^- \to \mu^+\mu^- + n\gamma}}{\dd\Omega_{1}} = 
\int\limits_{0}^{1}
\int\limits_{0}^{1}
\dd x_{1}\dd x_{2}{\cal D}(z_{1},s){\cal D}(z_{2},s)
\frac{\dd\tilde{\sigma}_{0}^{e^+e^- \to \mu^+\mu^-}(z_{1},z_{2})}
{\dd\Omega_{1}}
\bigl(1 + \frac{2\alpha}{\pi}\tilde{K}\bigr)\Theta({\mathrm{cuts}}) 
\nonumber \\ 
&& + \frac{\alpha}{\pi}\int\limits_{\Delta}^{1}\frac{\dd x_{1}}{x_{1}}
\biggl[\bigl(z_{1} + \frac{x_{1}^2}{2}\bigr) 
\ln\frac{\theta_{0}^{2}}{4} + \frac{x_{1}^2}{2}\biggr] 
\frac{\dd\tilde{\sigma}_{0}^{e^+e^- \to \mu^+\mu^-}(z_{1},1)}
{\dd\Omega_{1}}\Theta({\mathrm{cuts}}) 
\nonumber \\
&& + \frac{\alpha}{\pi}\int\limits_{\Delta}^{1}\frac{\dd x_{2}}{x_{2}}
\biggl[\bigl(z_{2} + \frac{x_{2}^2}{2}\bigr) 
\ln\frac{\theta_{0}^{2}}{4} + \frac{x_{2}^2}{2}\biggr]
\frac{\dd\tilde{\sigma}_{0}^{e^+e^- \to \mu^+\mu^-}(1,z_{2})}
{\dd\Omega_{1}}\Theta({\mathrm{cuts}}) 
\nonumber \\
&& + \frac{\alpha^3}{2\pi^2s^2}
\int\limits_{k^{0}>\Delta\varepsilon \atop \theta_{\gamma}>\theta_{0}}
R^{e^+e^- \to \mu^+\mu^-\gamma}_{\mathrm{hard}}
\frac{s\beta_{1}x\dd x\dd\Omega_{\gamma}}
{4[2 - x(1-\cos\psi/\beta_{1})]}\Theta({\mathrm{cuts}}) 
\nonumber \\
&& + \frac{2\alpha}{\pi}
\biggl[\frac{1 + \beta^2}{2\beta}\ln\frac{1 + \beta}{1 - \beta} 
- 1 + 2\ln\frac{1 - \beta c_{1}}{1 + \beta c_{1}}\biggr]
 \ln(\frac{\Delta\varepsilon}{\varepsilon})\cdot 
 \frac{\dd\tilde{\sigma}_{0}^{e^+e^- \to \mu^+\mu^-}(1,1)}
{\dd\Omega_{1}}\Theta({\mathrm{cuts}}),
\label{eq:master-mu}
\end{eqnarray}
where $\tilde{K} =
\pi^2/6-1/4+K_{\mathrm{even}}^{\mu}(\tilde {s}, \tilde {\theta_{1}}) + 
K_{\mathrm{odd}}^{\mu}(\tilde {s}, \tilde {\theta_{1}})$ and 
$\tilde {\theta_{1}}$ is a polar angle of negative muon in the c.m.frame.
The {\em master} formula drawn above  
provides within the scope of the discussion the intended cross section 
accuracy $\sim 0.2\%$. 
The integration limits of the first
term in Eq.\ref{eq:master-mu} were divided into two parts as for electrons. 
A two-fold integral splits into four separate contributions.
Those of them which describe one photon jet radiation are combined in a 
proper way with two {\em compensators} in the {\em master} formula. 
All other steps to construct the MC generator to simulate production 
of muon pair are similar to that as for electrons and can be found 
in~\cite{binp-70}.  

\begin{figure}[htbp]
  \begin{minipage}[t]{0.45\textwidth}
    \centering\includegraphics[width=0.95\textwidth]{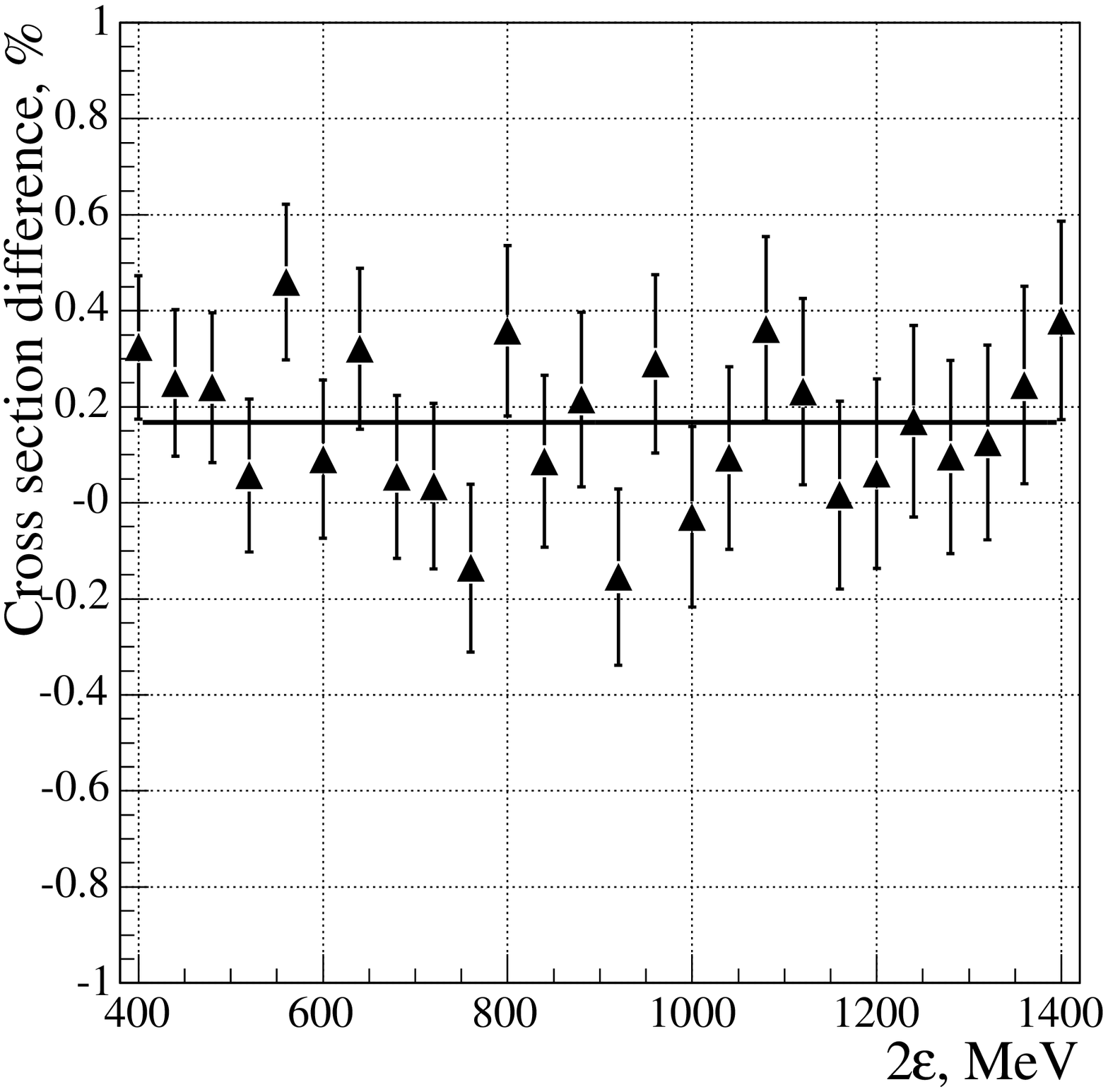}
    \caption{\label{fig:kkmcdiff} The relative difference between the 
      cross sections calculated by the MCGPJ code and KKMC {\it versus} the
      c.m.energy.}
  \end{minipage}
  \hfill
  \begin{minipage}[t]{0.45\textwidth}
    \centering\includegraphics[width=0.95\textwidth]{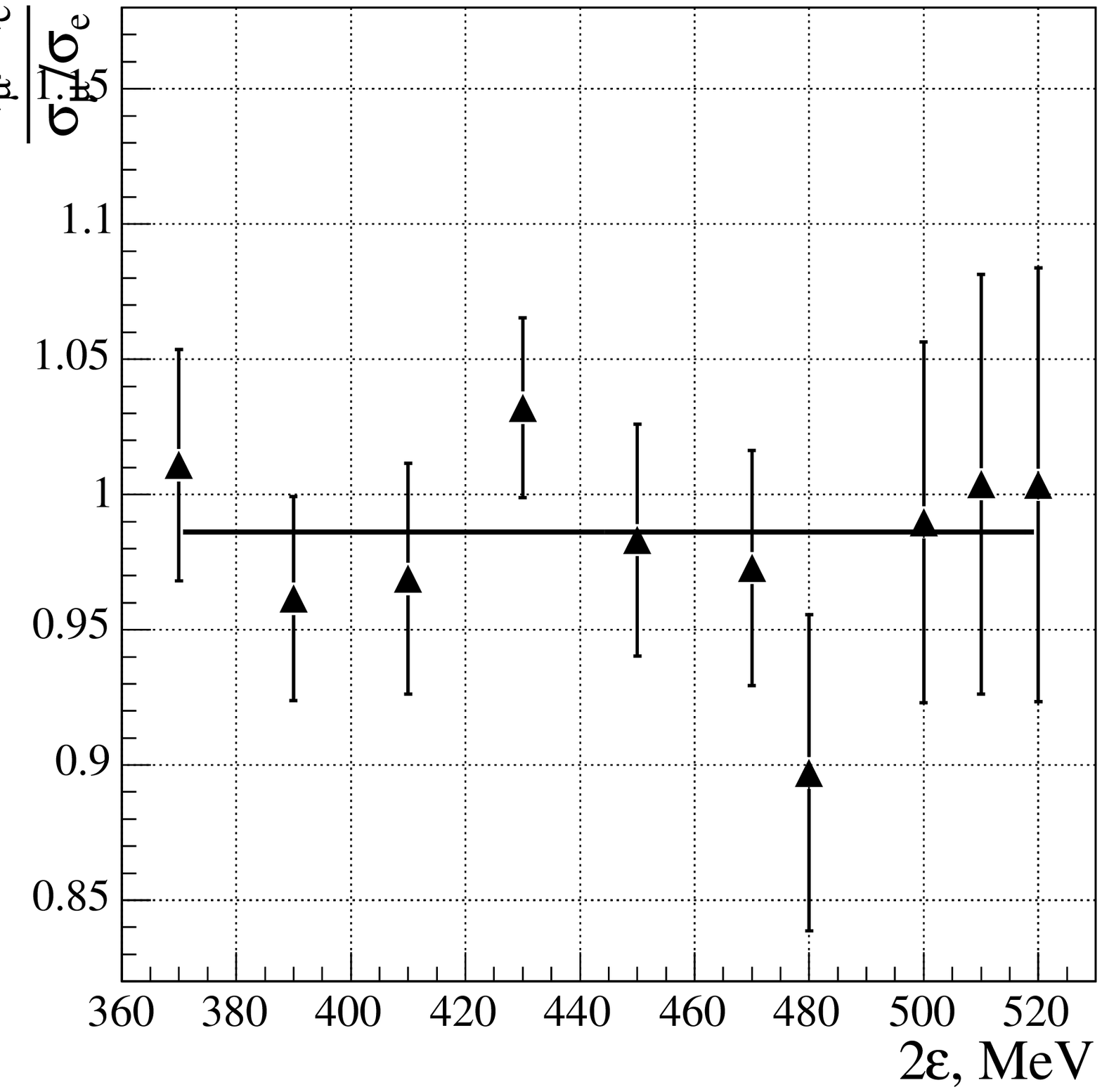}
    \caption{\label{fig:meratio} The ratio of the number of selected 
      muons to that of electrons 
 divided by the ratio of the corresponding theoretical cross sections.}
  \end{minipage}
\end{figure}

Numerical comparisons with the KKMC~\cite{KKMC} generator
have been performed. The
theoretical accuracy of the formulae on which KKMC is based is about  
$\sim 0.1\%$. The existing code in KKMC does not provide the correct 
description of vacuum polarization effects in the photon propagator 
at low energies, so they were switched off in both generators. 
The relative difference between cross sections calculated with 
the MCGPJ generator and KKMC in the VEPP-2M energy range is presented in
Fig.\ref{fig:kkmcdiff}. Good agreement at the level of our precision
$\pm 0.2\%$ is observed. 

In the low energy range, the momentum resolution of the CMD-2 detector 
is sufficient to distinguish pions, muons and electrons. Thus we have a  
direct way to compare the number of selected muons to that of 
electrons divided by the ratio of the theoretical cross sections, 
$\sigma (ee \to \mu\mu)/ \sigma (ee \to ee)$ and check thereby  
the theoretical precision of the formulae with RC from experiment. 
The results for this double ratio are presented in Fig.\ref{fig:meratio}.  
One can see that a deviation from unity of the double ratio  
does not exceed on average $1.4\%$ with statistical and systematic 
errors about $\sim 1.5\%$ and  $\sim 0.7\%$, respectively. 
The scarce experimental statistics in this energy range 
does not allow to make the comparison with better accuracy.

\section{Monte-Carlo generator for production of pion pairs}
The same ideas as for muons were applied here to construct the {\em master} 
formula to the processes $e^+e^-\to\pi^+\pi^-(n\gamma)$, $K^+K^-(n\gamma)$, 
$K_{S} K_{L}(n\gamma)$, assuming that pseudo scalar mesons are point-like 
objects. The enhanced contributions to the cross section, coming from 
the collinear region, are accounted for by means of SF formalism.
The one-loop virtual corrections, radiation of soft as well as one hard photon 
are taken into account in the first order of $\alpha$ exactly. 
The effects of the vacuum polarization are not included into the formulae 
presented below. In this case we deal with the so-called {\em dressed} cross 
section, when the dynamics of the pions strong interaction is encoded in the 
form factor properties. But, the Coulomb interaction in the final state 
should be included into RC to eliminate electromagnetic corrections. 

According to the papers~\cite{arkurpion, binp-70} the {\it boosted} Born cross
section is given by the expression
\begin{eqnarray}
\frac{\dd\tilde{\sigma}_{0}^{e^+e^- \to \pi^+\pi^-}(z_1,z_2)}{\dd\Omega_1} = 
\frac{\alpha^2}{4s}\frac{(Y_1^2-m_{\pi}^2/\eps^2)^{3/2}}{z_1^2z_2^2} 
  \frac{(1-c_{1}^2)|F_{\pi}(sz_1z_2)|^2}{z_1+z_2+(z_2-z_1)
(1-m^2_{\pi}/(\eps^2Y_1^2))^{-1/2}c_1}, 
\end{eqnarray}
where $z_{1,2}$ are the energy fractions of the electron and
positron after radiation of photon jets in the collinear region,
$|F_{\pi}(sz_{1}z_{2})|^2$ is a pion form factor squared, $c_1 =
\cos{\theta_1}$, $\theta_1$ is a polar angle between momentum of 
negative pion and electron beam direction.  
The energy fractions $Y_{1,2}$ of the final pions and the polar angle of
the positive pion, $\theta_2$, can be found from the same kinematic relations 
as for muons. 

\begin{figure}[htbp]
  \begin{minipage}[t]{0.45\textwidth}
    \centering\includegraphics[width=0.95\textwidth]
{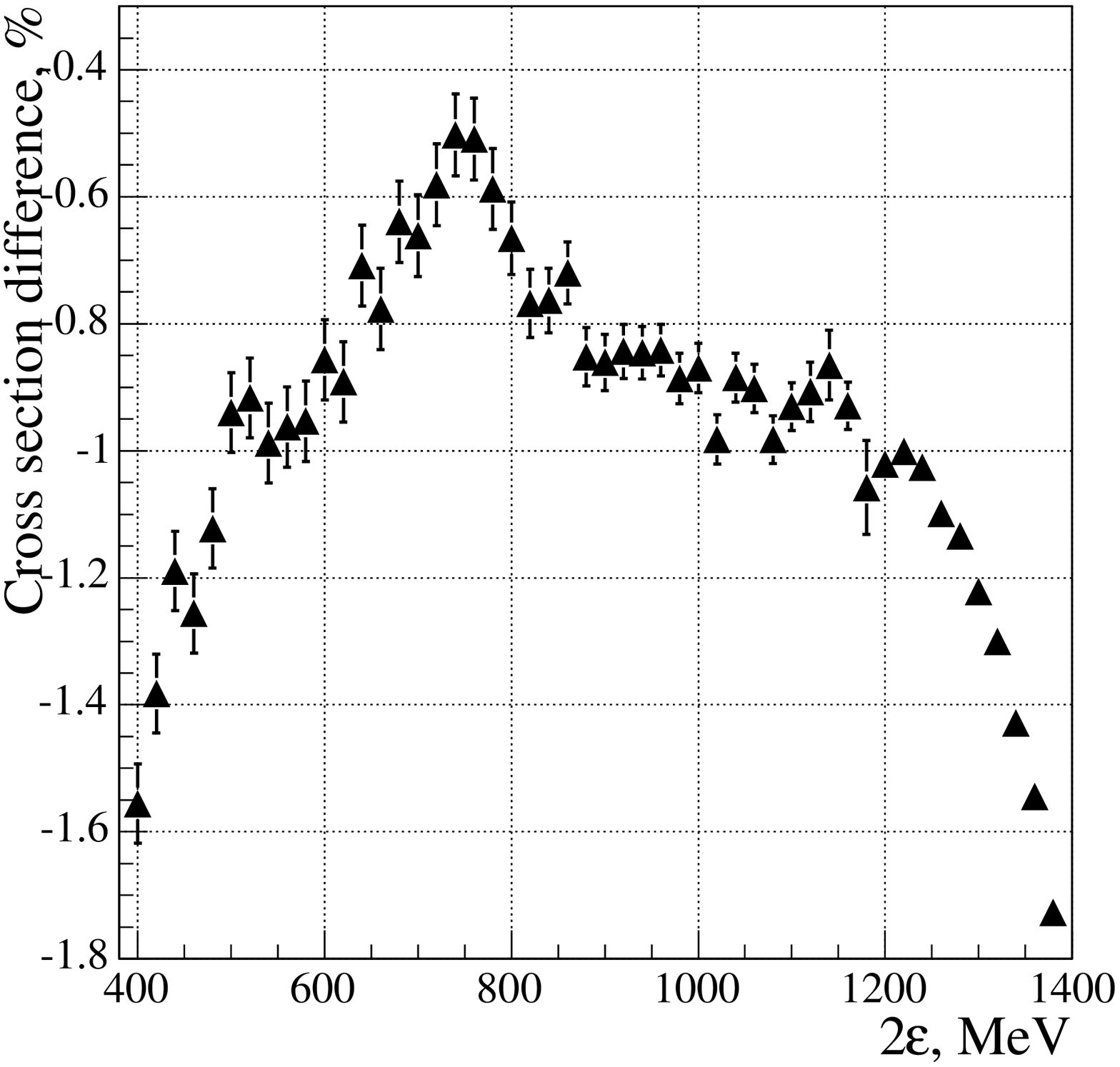}
    \caption{\label{fig:babapipidiff} The relative difference between  
the cross sections calculated by the MCGPJ code and BABAYAGA {\it versus} 
the beam energy.}
  \end{minipage}
  \hfill
  \begin{minipage}[t]{0.45\textwidth}
    \centering\includegraphics[width=0.99\textwidth]
{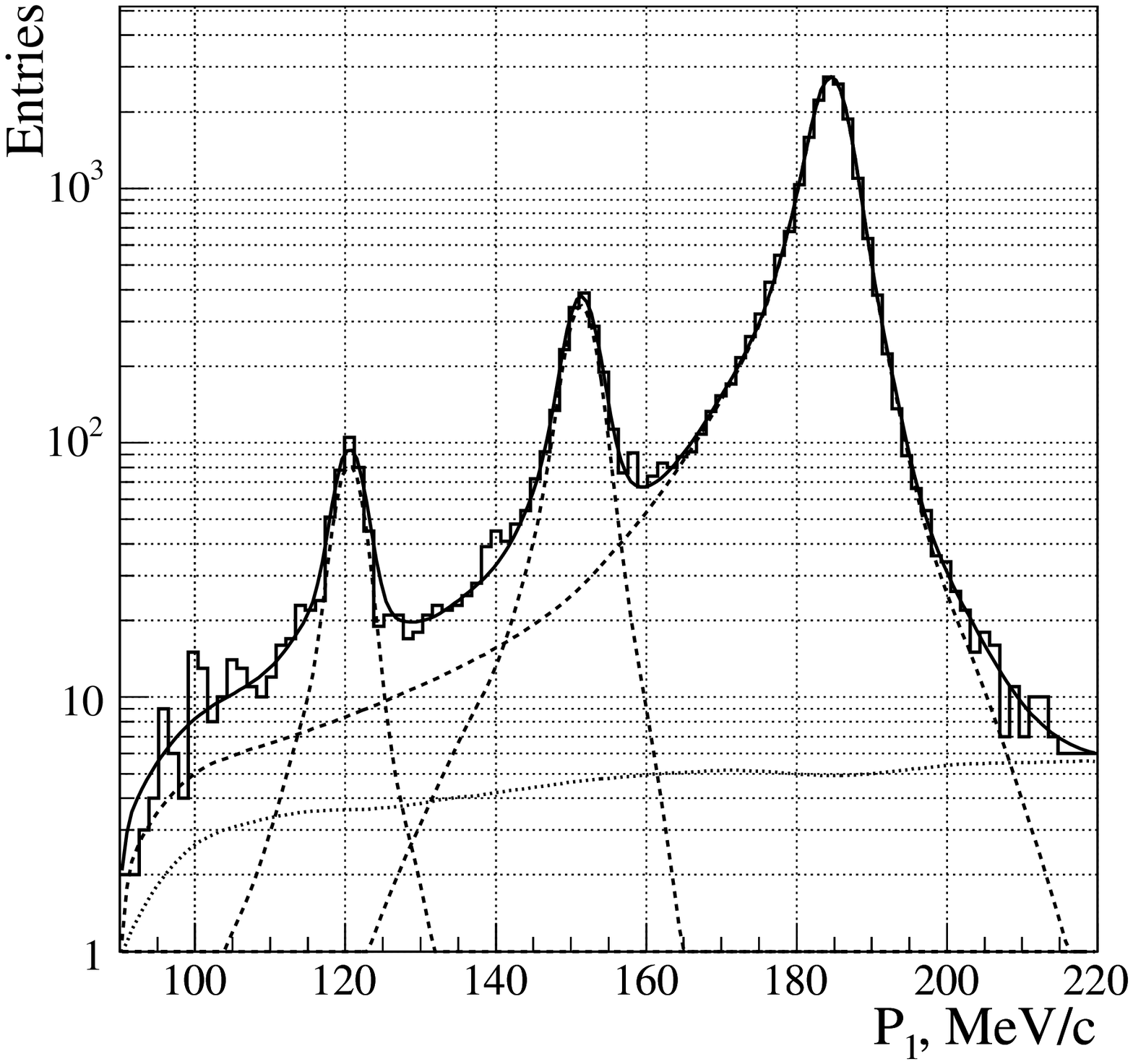}
\caption{\label{fig:sim-exp-390} The distribution of pion, 
muon and electron pairs 
as a function of the momentum. The upper curve represents a common fit, 
bottom curve - mainly cosmic ray background.}
  \end{minipage}
\end{figure}

The charge-even part of the cross section
due to radiation of soft and virtual photons~\cite{ch-even,Hoefer:2001mx} 
can be written in a convenient way as in~\cite{arkurpion}: 
\begin{eqnarray}
\frac{\dd\sigma^{S+V}_{\mathrm{even}}}{\dd\Omega_{1}} = 
\frac{\dd\sigma^{e^+e^- \to \pi^+\pi^-}_{0}(1,1)}{\dd\Omega_{1}}\cdot
\frac{2\alpha}{\pi} (A_{e} + A_{\pi}), 
\nonumber \\
A_{e} = (L - 1)\ln\frac{\Delta\varepsilon}{\varepsilon} + 
\frac{3}{4}(L - 1) + \frac{\pi^2}{6} - \frac{1}{4}, \hspace{3 mm} 
A_{\pi} = \biggl(\frac{1 + \beta^2}{2\beta}\ln\frac{1 + \beta}{1 - \beta} 
- 1 \biggr)\ln\frac{\Delta\varepsilon}{\varepsilon} + 
K_{\mathrm{even}}^{\pi}.
\end{eqnarray}
The expression for the quantity $K_{\mathrm{even}}^{\pi}$ can be found in 
~\cite{arkurpion,Hoefer:2001mx}. 
\begin{eqnarray}
K_{\mathrm{even}}^{\pi}=-1+\frac{1-\beta}{2\beta}\rho+\frac{2+\beta^2}
{\beta}\ln\frac{1+\beta}{2} 
+\frac{1+\beta^2}{2\beta}\left[\rho+\frac{\pi^2}{6} 
 + l_\beta\ln\frac{1+\beta^2}
{2\beta^2}+2\Li{\frac{1-\beta}{1+\beta}}\right].
\end{eqnarray}
The charge-odd part of the differential cross section   
is the interference result of the Born amplitude with those  
describing box-type diagrams and soft photons emission by electrons and
pions~\cite{ch-odd}. According to~\cite{arkurpion} the 
expression for the charge-odd part has the following form:
\begin{eqnarray}
\frac{\dd\sigma^{S+V}_{\mathrm{odd}}}{\dd\Omega_{1}} = 
\frac{\dd\sigma^{e^+e^- \to \pi^+\pi^-}_{0}(1,1)}{\dd\Omega_{1}}
\cdot\frac{2\alpha}{\pi}
\biggl(2\ln\frac{\Delta\varepsilon}{\varepsilon}\ln\frac{1 - \beta c_{1}}
{1 + \beta c_{1}} + K_{\mathrm{odd}}^{\pi}\biggr), 
\end{eqnarray} 
where $K_{\mathrm{odd}}^{\pi}$, in its turn, is equal to 
\begin{eqnarray} 
K_{\mathrm{odd}}^{\pi} &=& \frac{1}{2}l_-^2
- \mbox{Li}_2\left(\frac{1-2\beta c_{1} + \beta^2}{2(1-\beta c_{1})}\right)
+ \mbox{Li}_2\left(\frac{\beta^2(1-c_{1}^2)}{1-2\beta c_{1} + \beta^2}\right)
\\ \nonumber
&-& \int\limits_{0}^{1-\beta^2}\frac{\dd x}{x}f(x)\left(1
- \frac{x(1-2\beta c_{1} + \beta^2)}{(1-\beta c_{1})^2}\right)^{-\frac{1}{2}}
\nonumber \\
&+& \frac{1}{2\beta^2(1-c_{1}^2)}\Biggl\{\left[\frac{1}{2}l_-^2
- (L+l_-)L_-
+ \Li\left(\frac{1-\beta^2}{2(1-\beta c_{1})}\right) \right](1-\beta^2) 
\nonumber \\
&+& (1-\beta c_{1})\biggl[-l_-^2
- 2\Li\left(\frac{1-\beta^2}{2(1-\beta c_{1})}\right)
+ 2(L+l_-)L_-
\nonumber \\
&-& \frac{(1-\beta)^2}{2\beta}\left(\frac{1}{2}L^2 + \frac{\pi^2}{6}\right) 
+ \frac{1+\beta^2}{\beta}\biggl(
L\ln\frac{2}{1+\beta}
- \Li\left(-\frac{1-\beta}{1+\beta}\right)
\nonumber \\
&+& 2\Li\left(\frac{1-\beta}{2}\right) \biggr) \biggr] \Biggr\}
- (c_{1}\to -c_{1}), 
\\ \nonumber
f(x)&=&\biggl(\frac{1}{\sqrt{1-x}}-1\biggr)
\ln\frac{\sqrt{x}}{2} - \frac{1}{\sqrt{1-x}}
\ln\frac{1+\sqrt{1-x}}{2}\, ,
\\ \nonumber
l_-&=&\ln\frac{1-\beta c_{1}}{2}\, ,\qquad
L_-=\ln\biggl(1-\frac{1-\beta^2}{2(1-\beta c_{1})}\biggr).
\end{eqnarray}
The cross section of pion pair production accompanied by hard photon
emission can be presented in the following form~\cite{arkurpion}:  
\begin{eqnarray}
\frac{\dd\sigma^{e^+e^- \to \pi^+\pi^-\gamma}_{\mathrm{hard}}}{\dd\Omega_1} = 
\frac{\alpha^3}{32\pi^2s}R^{e^+e^- \to \pi^+\pi^-\gamma}_{\mathrm{hard}} 
\frac{s\beta_{1}x \dd x \dd\Omega_{\gamma}}{4[2 - x(1-\cos\psi/\beta_{1})]}.
\end{eqnarray} 
The quantity $R^{e^+e^- \to \pi^+\pi^-\gamma}_{\mathrm{hard}}$ 
contains the terms describing initial and final state radiation 
and their interference:
\begin{eqnarray}
&& R^{e^+e^- \to \pi^+\pi^-\gamma}_{\mathrm{hard}} =
R_{ee} + R_{\pi\pi} + R_{e\pi}, 
\nonumber \\
&& R_{ee} = |F_{\pi}(s_1)|^2\biggl\{ A\,\frac{4s}{\chi_-\chi_+}
      - \frac{8m_{e}^2}{s_1^2}\left(\frac{t_1u_1}{\chi_-^2}
      + \frac{tu}{\chi_+^2}\right)
+ \frac{8m_{e}^2m_{\pi}^2}{s_1}\left(\frac{1}{\chi_-^2}
    + \frac{1}{\chi_+^2}\right) + m_{\pi}^2\Delta_{s_1s_1}\biggr\}, 
\nonumber \\ &&
R_{\pi\pi}  = |F_{\pi}(s)|^2\biggl\{A\,\frac{4s_1}{\chi_-'\chi_+'}
      - \frac{8m_{\pi}^2}{s^2}\left(\frac{tu_1}{\chi_+'\!\!{}^2}
      + \frac{t_1u}{\chi_-'\!\!{}^2}\right) + m_{\pi}^2\Delta_{ss}\biggr\},  
\nonumber\\
&& R_{e\pi} = \Ree\;(F_{\pi}(s)F_{\pi}^*(s_1)) 
\biggl\{ 4A\biggl(\frac{u}{\chi_-\chi_+'} + \frac{u_1}{\chi_+\chi_-'}
- \frac{t}{\chi_-\chi_-'} - \frac{t_1}{\chi_+\chi_+'} \biggr) 
 + m_{\pi}^2\Delta_{ss_1} \biggr\}, 
\nonumber \\
&& A = \frac{tu+t_1u_1}{ss_1}\, ,\quad \Delta_{s_1s_1} = - \frac{4}{s_1^2}\,
    \frac{(t+u)^2+(t_1+u_1)^2}{\chi_+\chi_-}\, , 
\nonumber \\
&& \Delta_{ss} = \frac{2m_{\pi}^2(s-s_1)^2}{s(\chi_-'\chi_+')^2}
      + \frac{8}{s^2}(tt_1+uu_1-s^2-ss_1),
\nonumber \\
&& \Delta_{ss_1} = \frac{8}{ss_1}\biggl[
\frac{2(t_1-u)+u_1-t}{\chi_-'} + \frac{2(t-u_1)+u-t_1}{\chi_+'} 
\nonumber \\ && \quad
+ \frac{u_1+t_1-s}{2\chi_-}\biggl(\frac{u}{\chi_+'}
      - \frac{t}{\chi_-'}\biggr)
 + \frac{u+t-s}{2\chi_+}\biggl(\frac{u_1}{\chi_-'}
      - \frac{t_1}{\chi_+'}\biggr)\biggr].
\label{ree}
\end{eqnarray}
Mandelstam variables and $\chi_{\pm},$ $\chi'_{\pm}$ are defined as 
for electrons. The expression for the {\em master} formula, describing 
the process of pion pair production with two {\em compensators}, has 
a similar form as for muons and reads
\begin{eqnarray}  
&& \frac{\dd\sigma^{e^+e^- \to \pi^+\pi^- + n\gamma}}{\dd\Omega_{1}} = 
\int\limits_{0}^{1}
\int\limits_{0}^{1}
\dd x_{1}\dd x_{2}{\cal D}(z_{1},s){\cal D}(z_{2},s)
\frac{\dd\tilde{\sigma}_{0}^{e^+e^- \to \pi^+\pi^-}(z_{1},z_{2})}
{\dd\Omega_{1}}\biggl(1 + \frac{2\alpha}{\pi}\tilde{K}\biggr)
\Theta({\mathrm{cuts}}) 
\nonumber \\ && \quad
+ \frac{\alpha}{\pi}\int\limits_{\Delta}^{1}\frac{\dd x_{1}}{x_{1}}
\biggl[\bigl(z_{1} + \frac{x_{1}^2}{2}\bigr) 
\ln\frac{\theta_{0}^{2}}{4} + \frac{x_{1}^2}{2}\biggr] 
\frac{\dd\tilde{\sigma}_{0}^{e^+e^- \to \pi^+\pi^-}(z_{1},1)}
{\dd\Omega_{1}}\Theta({\mathrm{cuts}}) 
\nonumber \\ && \quad
+ \frac{\alpha}{\pi}\int\limits_{\Delta}^{1}\frac{\dd x_{2}}{x_{2}}
\biggl[\bigl(z_{2} + \frac{x_{2}^2}{2}\bigr) 
\ln\frac{\theta_{0}^{2}}{4} + \frac{x_{2}^2}{2}\biggr]
\frac{\dd\tilde{\sigma}_{0}^{e^+e^- \to \pi^+\pi^-}(1,z_{2})}
{\dd\Omega_{1}}\Theta({\mathrm{cuts}}) 
\nonumber \\ && \quad
+ \frac{\alpha^3}{32\pi^2s}
\int\limits_{k^{0}>\Delta\varepsilon \atop \theta_{\gamma}>\theta_{0}}
R^{e^+e^- \to \pi^+\pi^-\gamma}_{\mathrm{hard}}
\frac{s\beta_{1} x \dd x\dd\Omega_{\gamma}}
{4[2 - x(1-\cos\psi/\beta_{1})]}\Theta({\mathrm{cuts}})
\nonumber \\ && \quad
+ \frac{2\alpha}{\pi}
\biggl[\frac{1 + \beta^2}{2\beta}\ln\frac{1 + \beta}{1 - \beta} 
- 1 + 2\ln\frac{1 - \beta c_{1}}{1 + \beta c_{1}}\biggr]
 \ln\frac{\Delta\varepsilon}{\varepsilon} \cdot
 \frac{\dd\tilde{\sigma}_{0}^{e^+e^- \to \pi^+\pi^-}(1,1)}
{\dd\Omega_{1}}\Theta({\mathrm{cuts}}), 
\label{eq:master-pi}
\end{eqnarray}
where $\tilde{K} = \pi^2/6-1/4 + 
K_{\mathrm{even}}^{\pi}(\tilde {s}, \tilde {\theta_{1}}) + 
K_{\mathrm{odd}}^{\pi}(\tilde {s}, \tilde {\theta_{1}})$,
$\tilde {\theta_{1}}$ is a polar angle of negative pion  
in the center-of-mass system. 
As well as for muons, the integration limits with energy in the first 
term in Eq.\ref{eq:master-pi} were again divided in two parts. Two terms 
describing one photon jet radiation are merged with two {\em compensators}.   
Comparison with the BABAYAGA~\cite{babayaga} generator was performed.
The theoretical accuracy of the formulae, used in the BABAYAGA 
code, is about $\sim$1\%. The current version of the BABAYAGA code (3.5) 
doesn't include the FSR   
and so this term was removed from our code (just for comparison).  
The difference of the cross sections calculated by the MCGPJ generator and
BABAYAGA is shown in Fig.\ref{fig:babapipidiff} with the same
selection criteria as for Bhabha scattering events.  
A systematic shift between cross sections is on average about $1\%$  
in agreement with the BABAYAGA code precision, but for the lowest and 
highest energies the agreement becomes worse.   

The distributions of pion, muon and electron pairs as a function of 
momentum are presented in Fig.\ref{fig:sim-exp-390} at the c.m.s. 
energy of 390~MeV for experimental and simulated events.  
Momentum and angle resolutions, decays in flight,
interaction with the detector material and other important factors were
{\em smeared} with the parameters of simulated events to create events 
as close as possible to the {\em real} ones.
The histograms for each type of particles were fitted by two
Gaussian functions. Their relative weights and widths were the free
parameters of the fit. Perfect agreement between experiment and
simulation one can see. 

The enveloping curve describes pretty well the 
shape of histograms both at the {\em peaks} and at the
{\em tails}. It permits to determine the number of events inside each
histogram and to estimate the amount of the muons and electrons
under the pion peak and thereby to extract the systematic error
due to events separation procedure. 

The shape of histograms peaks of the simulated
events is not described well, if the MC generator, based on the
formulae in the first order in $\alpha$, is used. The shape of the  
histogram peak is mainly driven by the emission spectrum of soft photons and
the apparatus resolution. Thus, the number of events in 
the {\em tail} area is determined by the peak shape and hence, 
the approach with photon jet radiation is absolutely necessary.\\

The MC generator simulating  production of charged kaons   
is created similarly to that for pions. The pion mass
$m_{\pi}$ and form factor should be replaced in the 
above expressions by the kaon ones.
The cross section being multiplied by the exact Coulomb factor will 
interpolate the energy dependence of the cross section from  
the threshold production to the relativistic region.  
The exact expression for the Coulomb factor was obtained by 
Sommerfeld-Sakharov and reads 
\begin{eqnarray}
f(z) = \frac{z}{1 - \exp(-z)} - z/2, \qquad z = \frac{2\pi\alpha}{v}  
\end{eqnarray}
where $v$ is a relative velocity of koons. The term $z/2$ is subtracted 
because it is already included in the ${\mathcal O}(\alpha)$ corrections 
to final state. \\ 
 
The MC generator simulating production of neutral kaons   
is significantly simpler since there is no 
Coulomb interaction and photon emission in the final state. 
The quantity $R^{e^+e^- \to K_{L}K_{S}\gamma}_{\mathrm{hard}}$(\ref{ree}) 
consists of one term which describes the initial state radiation only and 
the value $\tilde{K}$ is equal to $\pi^2/6-1/4$. 

\section{Summary and concluding remarks}
The MC generator for the processes  
$e^+e^- \to e^+e^-$, $\mu^+\mu^-$, $\pi^+\pi^-$,
$K^+K^-$ and $K_{L}K_{S}$ is described in detail. 
An extended treatment of radiative corrections is implemented 
in the generator to get a high level of theoretical precision.
The current version of the program, Monte-Carlo Generator 
Photon Jets (MCGPJ), includes radiation effects 
in the first order in $\alpha$ exactly. 
The corrections deal with radiation of hard photon decomposed into the 
three parts which describe initial and final state radiation and their 
interference. All terms in the matrix elements 
which are proportional to the muon or pion mass squared are kept.
The enhanced contributions coming from the collinear 
region are accounted for by 
means of the SF formalism. As a result, the theoretical accuracy 
of the cross sections with RC is estimated to be at $\sim 0.2\%$ level. 
It is better by at least a factor of two compared to  
the accuracy $0.5 - 1\%$ achieved in earlier papers. Comparison with the 
well known codes BHWIDE, KKMC and BABAYAGA shows a satisfactory level of 
agreement for many distributions simulated by the generators.    

The shape of the distributions with acollinearity 
angles $\Delta\theta$ and $\Delta\phi$ agrees with the CMD-2 
experimental data.
The double ratio of the number of muon events to that of  
electrons divided by the ratio of the theoretical cross sections was 
found to be $0.986 \pm 0.014$. The deviation from unity is  
$-1.4 \pm 1.4\%$. This is the first direct comparison of the experimental 
cross sections with the theoretical calculation at the accuracy  
$\sim 1\%$. The comparison of momenta distributions  
at the lowest energy point  
shows that only the simulation with radiation of photon jets   
describes the experimental spectra pretty well.
Relying upon the above sketched review, the main conclusion is that 
theoretical predictions aiming at a $\mathcal O(0.1\%)$ precision 
must include contributions of both exact $\mathcal O(\alpha)$ terms 
and all higher order $\mathcal O(\alpha^n L^n)$ corrections. \\     

The theoretical uncertainties of the cross sections with RC are  
determined by the unaccounted higher order
corrections and they are estimated to be at $\sim 0.2\%$ level.
Below, the main sources of uncertainties in the current formulae are listed:
\begin{itemize}
\item 
The weak interaction contributions are omitted in our approach. 
The numerical estimations show that for energies $2\varepsilon < 3$~GeV 
these contributions do not exceed $0.1\%$.
\item 
A part of the second order next-to-leading radiative corrections
proportional to $(\alpha/\pi)^2  L \sim 10^{-4}$ were omitted.
Among these contributions are: the effect due to double photon
emission (one inside and one outside of the narrow cone); emission of 
soft or virtual photon simultaneously with radiation of one hard photon 
at large angles. Even if we assume 
that a coefficient in front of these terms will be of order of ten, 
their contribution can not exceed $\sim 0.1\%$. 
\item 
The next source of uncertainties is related to the calculation 
of the hadronic vacuum polarization contribution to the photon 
propagator. Numerical estimations show that a systematic error 
of hadronic cross sections of about 1\% changes the  
cross section by less than $\sim 0.04\%$.
\item 
The uncertainty of about 0.1\% is related to  
the theoretical models which are used to describe the energy 
dependence of the hadronic cross sections.
\item 
In Ref.~\cite{sm-vol} it was concluded that a combined effect of  
all the parametrically enhanced $\mathcal O (\alpha^2)$ corrections  
can be numerically limited by $2.0\times 10^{-4}$ for near threshold 
production. The magnitude of this contribution slowly decreases with 
the final particle velocity $\beta$ and therefore these corrections are 
beyond the intended accuracy. 
\item 
The last source of uncertainty is mainly driven by the 
collinear kinematic approximation. Several terms proportional to   
$(\alpha/\pi)\theta_{0}^2$ and $(\alpha/\pi)(1/\gamma \theta_{0})^2$ 
were omitted. Numerical estimations show that the contribution of 
these factors is about $\sim 0.1\%$. 
\end{itemize}
Considering the uncertainty sources mentioned above as independent, 
we can conclude that the total systematic error of the cross 
sections with RC is less than 0.2\%. An indirect confirmation 
of the correct evaluation of the accuracy is the comparison of
cross sections with RC calculated in the first order of $\alpha$ only. 
The corresponding difference does not exceed $0.2\%$. From that follows 
that higher order enhanced contributions, coming from collinear regions 
with emission of two and more photons, contribute to the cross section   
by amount of $\sim 0.2\%$ only for our selection criteria.        
Since the accuracy of this contribution is certainly known better than 
$100\%$, the systematic theoretical uncertainty for 
the cross sections with RC is $\sim 0.2\%$  \\

The authors are grateful to V.S.~Fadin, A.V.~Bogdan, A.I.~Milstein and 
G.N.~Shestakov, to all members of the CMD-2 collaboration and particularly 
to S.I.~Eidelman and B.I.Khazin for fruitful and 
useful discussions. We are also grateful to S.~Jadach and W.~Placzek 
for their help in runing the BHWIDE and KKMC codes, G.~Montagna and 
C.M.~Carloni Calame for the useful collaboration on the BABAYAGA code. 

This work is supported in part by the grants: RFBR-99-02-17053,
RFBR-99-02-17119, RFBR-03-02-17077 and INTAS 96-0624.


\begin{thebibliography}{99}
\bibitem{e821} G.V.~Bennett {\it et al.}, Phys. Rev. Lett. {\bf 92} (2004) 161802.
\bibitem{jeger} S.~Eidelman and F.~Jegerlehner, Z. Physik C {\bf 67} (1995) 585,\\
M.~Davier {\it et al.}, Eur. Phys. J. C {\bf 27} (2003) 497;\\
M.~Davier, Nucl. Phys. B (Proc. Suppl.){\bf 131} (2003) 192;\\
T.~Teubner, Nucl. Phys. B (Proc. Suppl.){\bf 131} (2003) 201;\\  
M.~Davier, S.~Eidelman, A.~Hocker and Z.~Zhang, Eur. Phys. J. C 31 (2003) 503.  
\bibitem{cmd2} R.R.~Akhmetshin {\it et al.}, Phys. Lett. B {\bf 89} (2002) 161;\\
R.R.~Akhmetshin {\it et al.}, Nucl. Phys. B (Proc. Suppl.){\bf 131} (2003) 3.
\bibitem{integr} B.~Lautrup, A.~Peterman and E. de Rafael, 
Phys. Rep. {\bf 3} (1972) 193.
\bibitem{e969} (g-2) Collabaration, Proposal E969, BNL (2004).
\bibitem{fadkur} E.A.~Kuraev and V.S.~Fadin, Sov. J. Nucl. Phys.{\bf 41} (1985) 466;\\
 S.~Yadach, M.~Skrzypek and B.F.L.~Ward, Phys.Lett. B {\bf 257} (1991) 173; \\
 M.~Skrzypek, Acta Phys. Polon, B {\bf 23} (1992) 135. \\
 M.~Cacciari, A.~Deandrea, G.~Montagna and O.~Nicrosini, Europhys. Lett. 
{\bf 17} (1992) 123;\\
  A.B.~Arbuzov {\it et al.}, JETP {\bf 81} (1995) 638;\\
   A.B.~Arbuzov {\it et al.}, Nucl. Phys. B {\bf 485} (1997) 457;
\bibitem{berends} F.A.~Berends {\it et al.}, Nucl. Phys. B {\bf 122} (1977) 485;\\
F.A.~Berends {\it et al.}, Nucl. Phys. B {\bf 57} (1973) 381;\\
 Erratum-ibid. B {\bf 75} (1974) 546;\\
   F.A.~Berends, K.J.F~Gaemers and R.~Gastmans, Nucl. Phys. B {\bf 68} (1974) 541;\\ 
    F.A.~Berends and R.~Kleiss, Nucl. Phys. B {\bf 228} (1983) 537;\\
     S.I.~Eidelman and E.A~Kuraev, Phys. Lett. B {\bf 80} (1978) 94.
\bibitem{arkurlept} A.B.~Arbuzov {\it et al.}, JHEP {\bf 10} (1997) 001.
\bibitem{arkurpion} A.B.~Arbuzov {\it et al.}, JHEP {\bf 10} (1997) 006.
\bibitem{tsai} Z.~Jakubowski {\it et al.}, Z. Phys. C {\bf 40} (1988) 49; \\
               Y.S.~Tsay, SLAC - PUB - 1515 (1973).
\bibitem{BFKh} V.N.~Bayer, V.S.~Fadin, V.A.~Khoze, Nucl. Phys. B {\bf 65} (1973) 381.
\bibitem{bhwide} S.~Jadach, W.~Placzek, B.F.L.~Ward, Phys. Lett. B {\bf 390} 
(1997) 298.
\bibitem{csmumu} V.N.~Bayer and V.A.~Khoze, JETP, vol.{\bf 48} (1965) 946;\\ 
                 S.I.~Eidelman, E.A.~Kuraev, V.S.~Panin,
                 Nucl. Phys. B {\bf 148} (1979) 245; \\
                 F.A.~Berends {\it et al.}, Phys.\ Lett.\ B {\bf 103}, (1981) 124.\\
                 E.A.~Kuraev, G.V.~Meledin, Nucl. Phys. B {\bf 122} (1977) 485.
\bibitem{khrip} I.B.~Khriplovich, Sov. J. Yad. Fiz, {\bf 17} (1973) 576; \\ 
                D.A.~Dicus, Phys. Rev. D {\bf 8} (1973) 890;\\ 
                F.A.~Berends and R.~Kleiss, Nucl. Phys. B {\bf 177} (1981) 237.
\bibitem{KKMC} S.~Jadach {\it et al.}, Comp. Phys. Commun. {\bf 130} (2000) 260.

\bibitem{binp-70} A.B.~Arbuzov, G,V.~Fedotovich {\it et al.}, Preprint, BINP 2004-70

\bibitem{ch-even} R.W.~Brown and K.O.~Mikaelian, Lett. Nuovo Cim. {\bf 10} (1974) 305.
\bibitem{Hoefer:2001mx}
A.~Hoefer, J.~Gluza and F.~Jegerlehner,
Eur.\ Phys.\ J.\ C {\bf 24} (2002) 51.
\bibitem{ch-odd} V.N.~Bayer, VIII Winter School LINP, v.II (1973) 164; \\
                 O.P.~Sushkov, Sov. J. Yad. Fiz. {\bf 22} (1975) 868.
\bibitem{babayaga} C.M.~Carloni Calame {\it et al.}, Nucl. Phys {\bf B584} 
                   (2000) 459; \\
                   C.M.~Carloni Calame, Phys. Lett. {\bf B520} (2001) 16. 

\bibitem{sm-vol} M.B.~Voloshin, Preprint, TPI-MINN-02/49-T, arXiv:hep-ph/021207, v1.
\end{thebibliography}
\end{document}